\documentclass[preprint,12pt]{imsart}

\usepackage{amssymb,amsmath}

\RequirePackage[OT1]{fontenc}
\RequirePackage{amsthm,amsmath}
\RequirePackage[numbers]{natbib}
\RequirePackage[colorlinks,citecolor=blue,urlcolor=blue]{hyperref}

\usepackage{amscd,amsfonts,amssymb,amsmath,latexsym,array,hhline,xcolor,graphicx}
\usepackage{float}
\def \a{\alpha}
\def \b{\beta}

\def \l{\lambda}

\def \p{\pi}

\def \s{\sigma}

\def \o{\omega}
\def \O{\Omega}
\def \e{\varepsilon}

\newcommand\F{\mbox{I\kern-2pt F}}

\newcommand\cF{{\cal F}}
\newcommand\cG{{\cal G}}

\newcommand\cB{{\cal B}}
\newcommand\cH{{\cal H}}
\newcommand\cK{{\cal K}}

\newcommand\cX{{\cal X}}

\newcommand\cR{{\cal R}}

\newcommand\cP{{\cal P}}

\newcommand\R{{\mathbb{R}}}

\newcommand\E{\mathbb{E}}

\newtheorem{theo}{Theorem}[section]
\newtheorem{prop}[theo]{Proposition}
\newtheorem{lemm}[theo]{Lemma}
\newtheorem{defi}[theo]{Definition}
\newtheorem{ex}[theo]{Example}

\newtheorem{coro}[theo]{Corollary}
\newtheorem{rem}[theo]{Remark}
\newcommand\fdem{$\Box$}
\newcommand\beq{\begin{equation}}
\newcommand\eeq{\end{equation}}
\newcommand\bea{\begin{eqnarray}}
\newcommand\eea{\end{eqnarray}}
\newcommand\bean{\begin{eqnarray*}}
\newcommand\eean{\end{eqnarray*}}

\newcommand\bP{{\bf {\rm P}}}

\newcommand{\esssup}[1][\cH]{\mathrm{ess\,sup}_{#1}}
\newcommand{\essinf}[1][\cH]{\mathrm{ess\,inf}_{#1}}
\newcommand{\supp}[1][\cH]{\mathrm{ supp\,}_{#1}}

\begin{document}

\begin{frontmatter}



\title{Pricing without martingale measure}

\author[A1]{Julien Baptiste,}
\author[A2]{Laurence Carassus,} 
\author[A1,A3]{Emmanuel L\'epinette,}

 \address[A1]{ Paris Dauphine university, PSL research university, Ceremade,  CNRS, UMR, Place du Mar\'echal De Lattre De Tassigny, 75775 Paris cedex 16, France.\\
 Email: baptiste@ceremade.dauphine.fr, emmanuel.lepinette@ceremade.dauphine.fr}

\address[A2]{L\'eonard de Vinci P\^ole Universitaire, Research Center, 92 916 Paris La D\'efense, France \\
and LMR, universit\'e de Reims-Champagne Ardenne, France \\
Email: laurence.carassus@devinci.fr}

 \address[A3]{Gosaef,  Facult\'e des Sciences de Tunis, 2092 Manar II-Tunis, Tunisia.}

\begin{abstract}
For several decades, the no-arbitrage (NA) condition and the martingale measures have played a major role in the financial asset's pricing theory.  We propose a new approach for estimating the super-replication cost based on convex duality instead of martingale measures duality~: Our prices will be expressed using Fenchel conjugate and bi-conjugate. The super-hedging problem
leads  endogenously to  a weak condition of NA called Absence of Immediate Profit (AIP).
We propose several characterizations of AIP and  study the relation with the classical notions of no-arbitrage.
We also give some promising numerical illustrations.
\end{abstract}



\end{frontmatter}



\section{Introduction}
The problem of giving a fair price to a financial asset $G$ is central in the economic and financial theory. A selling price should be an amount which is enough to initiate a hedging strategy for $G$, i.e. a strategy whose value at maturity is always above $G$. It seems also natural to ask for the infimum of such amount. This is the so called   super-replication price and  it has been
introduced in the binomial setup for transaction costs by \cite{BLPS}. Characterizing and computing the super-replication price has become one of the central issue in mathematical finance theory. Until now it was intimately related to the No-Arbitrage (NA) condition. This condition asserts that starting from a zero wealth it is not possible to reach a positive one (non negative almost surely and strictly positive with strictly positive probability measure). Characterizing   the NA condition or, more generally, the  No Free Lunch condition leads to the Fundamental Theorem of Asset Pricing (FTAP in short). This theorem proves the equivalence  between  those absence of arbitrage conditions  and the existence of  equivalent risk-neutral probability measures (also called martingale measures or pricing measures) which are equivalent probability measures under which the  (discounted) asset price process is a martingale. This was initially formalised in \citep{HK79}, \citep{HP81} and \citep{K81} while in \citep{DMW}   the FTAP is formulated  in a general discrete-time setting under the NA condition. The literature on the subject is huge  and we refer to \citep{DelSch05} and \citep{KS} for a general overview.
Under the NA condition, the super-replication price of $G$ is equal to the supremum of the (discounted) expectation of $G$ computed under the  risk-neutral probability measures. This is  the so called dual formulation of the super-replication price or  superhedging theorem. We refer to  \cite{Schal99} and \cite{FK97} and the references therein.

In this paper, a super-hedging or super-replicating price is the initial value of some super-hedging strategy. We do not postulate any assumption on the financial market and analyze from scratch the  set of super-hedging prices and its infimum value, which will be called  the infimum super-hedging cost.
Under mild assumptions, we show that the one-step set of super-hedging prices can be expressed using Fenchel-Legendre conjugate and the infimum super-replication cost is obtained by the Fenchel-Legendre biconjugate. So,  we use here the convex duality instead of the usual financial duality based on martingale measures under the NA condition.
To do so, we use the notion of conditional essential supremum. Using measurable selection techniques, we show that
the conditional essential supremum of a function of $Y$ is equal to the usual supremum of the function evaluated on a random set, the conditional support of $Y$  (see Proposition \ref{lemma-essup-h(X)}).
The pricing formula that we obtain (see \eqref{prixavecAIP})
shows that, if  the initial stock price  $y$ does not belong to the convex hull of the conditional  support of the stock value at the end of the period $Y,$  then the super-hedging cost is equal to $-\infty$. To exclude this possibility we postulate
the condition of Absence of Immediate Profit  (AIP).
The AIP is an endogenous condition for pricing and is indeed very weak~: If the initial information is trivial, a one period immediate profit is a strategy which starts from 0 and leads to a deterministic strictly positive gain at time 1.
We propose several characterization of the AIP condition. 
In particular we show that AIP is equivalent to the non-negativity of the super-hedging prices of any fixed call option.
We also discuss in details the link between AIP and the others no-arbitrage conditions as 
the no-arbitrage of first and second type and the no-riskless arbitrage of \cite{Ing} and the No Unbounded Profit with Bounded Risk of \cite{KK}. None of the conditions is equivalent to AIP, the closest being the no-riskless arbitrage. 
Under AIP condition, we show that the one-step infimum super-hedging cost is the concave envelop of the payoff relatively to the convex envelop of the conditional support.
Fenchel-Legendre duality have already been used to obtain a dual representation of the super-replication price thanks to deflators (see \cite[Exemple 4.2]{PenMOR} and \cite[Theorem 10 and Corollary 15]{PenMF}).  In \cite[Theorem 10]{PenMF}  the result is shown under the assumption that the set
of claims that can be super-replicate from $0$
is closed, which holds true under NA.   Our approach is different as we do not postulate any assumption on the market and, actually, we do not seek for a dual representation of the (minimal) super-hedging price.

We then consider the multiple-period framework. We show that the global AIP condition and the local ones are equivalent.
We study the link between AIP, NA and the absence of  weak immediate profit  (AWIP)  conditions. We show that the AIP condition is the weakest-one and we also provide conditions for the equivalence between  the AIP and the AWIP  conditions, as well as characterization through absolutely continuous martingale measure.

We then focus on a particular, but still general setup, where we propose a recursive scheme for the computation of  the super-hedging prices of a convex option. We obtain the same computation scheme as in \cite{CGT} and \cite{CV} but here it is obtained by only  assuming  AIP instead of the stronger NA condition.
We also give  some numerical illustrations. We  calibrate  historical data of the  french index CAC $40$ to our model and  implement our super-hedging strategy for a call option. Our procedure is somehow model free as it is only based on statistical estimations.


The paper is organized as follows. In Section \ref{secone}, we study the one-period framework while in Section \ref{secmulti} we study the multi-period one.  Section \ref{seexpli} proposes an explicit pricing for a convex payoff  and numerical experiments.


In the remaining of this introduction we present our framework and notations. Let $(\Omega,(\cF_t)_{t \in \{0,\ldots,T\}}\cF_T,P)$ be a complete filtered probability space,  where $T$ is the time horizon.
For any $\sigma$-algebra $\cH$ and any $k \geq 1$,  we denote by $L^0(\R^k,\cH)$ the set of $\cH$-measurable and $\R^k$-valued random variables.
We consider   a   non-negative process
$S:=\left\{S_{t},\ t \in \{0,\ldots,T\},\right\}$ such that  $S_{t} \in L^0(\R^d,\cF_t)$ for all $t \in \{0,\ldots,T\}$. The vector $S_t$ represents the  price at time $t$ of the $d$ risky assets in the
financial market in consideration.
Trading
strategies are given by a   process $\theta:=\{ \theta_{t}, t \in \{0,\ldots,T-1\},\}$ such that  $\theta_{t} \in L^0(\R^d,\cF_{t})$ for all $t\in \{0,\ldots,T-1\},$. The vector $\theta_{t}$ represents the
investor's holding in   the $d$ risky assets between time $t$ and time $t+1$.
We assume that trading is self-financing and that the riskless asset's price is a constant equal to $1$. The value at time $t$ of a portfolio $\theta$ starting from
initial capital $x\in\mathbb{R}$ is then given by
$$
V^{x,\theta}_t=x+\sum_{u=1}^t  \theta_{u-1} \Delta S_u,
$$
where $\Delta S_u=S_u-S_{u-1}$ for $u \geq 1$ and $xy$ is the scalar product  of $x$ and $y$.

\section{The one-period framework}
\label{secone}
Let $\cH$ and $\cF$ be two complete sub-$\sigma$-algebras of $\cF_T$ such that $\cH \subseteq \cF$ and which represent respectively the initial and the final information. Let  $y\in L^0(\R^d,\cH)$   and $Y\in L^0(\R^d,\cF)$ be  two non-negative\footnote{For ease of notation, we assume that $y(\o) \geq 0$ and $Y(\o) \geq 0$ for all $\o \in \Omega$.} random variables.   They represents the initial and the final prices of the $d$ risky assets.
Finally, we introduce $g: \Omega \times \R\to \R$ and the associated derivative $g(Y)$, where $g(Y):\o \to g(Y)(\o)=g(\omega,Y(\omega)).$ \\
The objective of the section  is to obtain under suitable assumptions on $g$ a characterization of   $\cP(g)$,  the one-step set of super-hedging (or super-replicating) prices of $g(Y)$ and of its infimum value. The setting will be applied in Section \ref{secmulti} with the choices $\cH=\cF_t$, $\cF=\cF_{t+1}$, $Y=S_{t+1}$ and $y=S_t$.
\begin{defi}
The set $\cP(g)$ of super-hedging prices of the contingent claim $g(Y)$ consists in  the initial values of super-hedging strategies $\theta$:
$$\cP(g)=\{x\in  L^0(\R,\cH),  \exists\, \theta\in L^0(\R^d,\cH),\;   x+\theta (Y -y)\ge g(Y) \, {\rm a.s.}\}.$$
The infimum super-hedging cost of $g(Y)$ is defined by $p(g):=\essinf[\cH]\cP(g)$.
\end{defi}
The notions of conditional essential infimum $\essinf[\cH]$ and conditional essential supremum  $\esssup[\cH]$ are at the heart of this study and will be defined in Proposition \ref{Essup}  below. We will also use the conditional support of $Y$
${\rm supp}_{\cH}Y$  which is introduced in Definition \ref{DefD} below.
In Section \ref{secgreve}, we derive the characterization of $\cP(g)$ and $p(g)$ from the following steps~:
\begin{enumerate}
\item Observe that the set of super-hedging prices can be rewritten using a conditional essential supremum (see \eqref{eqpg}).
\item Show that under mild conditions the conditional essential supremum of a function of $Y$ is equal to the usual supremum of the function evaluated on the random set ${\rm supp}_{\cH}Y$ (see Proposition \ref{lemma-essup-h(X)}).
\item Recognize that a super-hedging price can be written using a Fenchel-Legendre conjugate (see \eqref{eqf*}).
\item Take the essential infimum of the set of super-hedging prices and go through the three first steps to recognize the Fenchel-Legendre biconjugate (see \eqref{eqbiconj}).
\item Use the classical convex biconjugate theorem (see  \eqref{eqpasiconcon} and Proposition \ref{fench}) to evaluate the infimum super-hedging cost.
\end{enumerate}
With this pricing formula in hand (see \eqref{prixavecAIP}), the condition of Absence of Immediate Profit  (AIP) appears endogenously. In Section \ref{secAIP}, we develop the concept of AIP and  propose several characterization of the AIP condition and compare it with the classical No Arbitrage NA condition.

\subsection{Conditional support and conditional essential infimum}
\label{consupessinf}
This section is the toolbox of the paper.
We recall some results and notations that will be used without  further references in the rest of the paper.
 Let  $h: \Omega \times \R^d \to \R$.
The effective domain of $h(\o,\cdot)$ is defined by
$${\rm dom \,} h(\o,\cdot):=\{x \in \R^d, \, h(\o,x)< \infty\}$$ and $h(\o,\cdot)$ is proper if dom $h(\o,\cdot) \neq \emptyset$ and $h(\o,x)>- \infty$  for all $x \in \R^d$.
Next,   if $h$  is $ \cH$-normal integrand (see Definition 14.27 in \cite{rw}) then $h$ is $\cH \otimes \cB(\R^d)$-measurable and is lower semi-continuous (l.s.c. in the sequel, see \cite[Definition 1.5]{rw}) in $x$ and the converse holds true if $\cH$ is complete for some measure, see \cite[Corollary 14.34]{rw}.
Note that, if $Z \in L^0(\R^d,\cH)$ and $h$ is $\cH \otimes \cB(\R^d)$-measurable, then
$h(Z) \in L^0(\R^d,\cH)$.\\
A random set $\cK : \Omega \twoheadrightarrow \mathbb{R}^{d}$ is said $\cH$-measurable if for all open set $O$ of $\R^d$, the subset $\{\o \in \O, \, O \cap \cK(\o) \neq \emptyset \} \in \cH$.
If $\cK$ is a $\cH$-measurable and closed-valued random set of $\R^d$, then $\cK$ admits a Castaing representation $(\eta_n)_{n\in \mathbb{N}}$ (see Theorem 14.5 in \cite{rw}). This means that
$\cK(\o)={\rm cl}\{\eta_n(\o),\,n\in \mathbf{N}\}$ for all $\o \in {\rm dom \, } \cK =\{\o \in \Omega, \, \cK(\o)\cap \R^d \neq \emptyset\}$,  where the closure is taken in $\R^d$.

First, we introduce  the conditional support of $X \in L^0(\R^d,\cF)$ with respect to $\cH$.
\begin{defi}
\label{DefD}
Let $\mu$ be a $\cH$-stochastic kernel (i.e. for all $\o \in \Omega$, $\mu(\cdot,\o)$ is a probability on $\cB(\R^d)$ and  $\mu(A,\cdot)$ is $\cH$-measurable for all $A \in \cB(\R^d)$). We define the random set  ${D}_{\mu} : \Omega \twoheadrightarrow \mathbb{R}^{d}$ by~:
\begin{align}
\label{defd1}
{D}_{\mu}(\o):=\bigcap \left\{ A \subset \mathbb{R}^{d},\; \mbox{closed}, \; \mu(A,\o)=1\right\}.
\end{align}
For $\o \in {\O}$, ${D}_{\mu}(\o) \subset \mathbb{R}^d$ is called the support of $\mu(\cdot, \o)$. \\
Let $X\in L^0(\R^d,\cF)$, we denote by $\supp[\cH]X$ the set defined in \eqref{defd1} when $\mu(A,\o)=P(X \in A | \mathcal{H})(\o)$ is a regular version of the conditional law of $X$ knowing $\cH$. The random set
 $\supp[\cH]X$ is called the conditional support of $X$ with respect to $\cH$.
\end{defi}
\begin{rem}
\label{remsupp}
When $\cH$ is the trivial sigma-algebra, $\supp[\cH]X$ is just  the usual support of $X$   (see p441 of   \cite{ab}).
Theorems 12.7  and 12.14 of \cite{ab} show  that
$P(X \in . | \mathcal{H})$ admits a unique support $\supp[\cH]X\subset \mathbb{R}^d$ such that we have   $P(X \in \supp[\cH]X | \mathcal{H})=1$ a.s. i.e.  ${\rm supp}_{\cH}X$ is a.s. non-empty.\\
For simplicity we will assumed that $Y(\o) \in \supp[\cH]Y (\o)$ for all $\o \in \Omega$.
Moreover, as  $0\leq Y < \infty$, ${\rm  Dom \; supp \; }_{\cH}Y=\Omega$.
\end{rem}
\begin{lemm}
\label{Dmeasurability}
$D_{\mu}$ is non-empty, closed-valued, $\mathcal{H}$-measurable and  graph-measurable random set (i.e.  $Graph(D_{\mu}) \in \mathcal{H} \otimes \mathcal{B}(\mathbb{R}^{d})$).
\end{lemm}
{\sl Proof.}
It is clear from \eqref{defd1} that, for all $\o \in \O$,  $D_{\mu}(\o)$ is a non-empty and closed subset of $\mathbb{R}^{d}$. We show that $D_{\mu}$ is  $\mathcal{H}$-measurable. Let $O$ be a fixed open set in $\mathbb{R}^{d}$ and
$
\mu_{O}: \omega \in \O \mapsto \mu_{O}(\o) :=  \mu(O,\o).
$
As $\mu$ is a stochastic kernel,  $\mu_{O}$ is $\mathcal{H}$-measurable.
By definition of  $D_{\mu}(\o)$ we get that $\{\o \in \Omega,\; {D}_{\mu}(\o) \cap O \neq \emptyset \} =\{\o \in \Omega, \; \mu_{O}(\omega)>0\} \in \mathcal{H},$
and $D_{\mu}$ is $\mathcal{H}$-measurable.
Now using Theorem 14.8 of \cite{rw}, $Graph(D_{\mu}) \in \mathcal{H} \otimes \mathcal{B}(\mathbb{R}^{d})$ (recall that $D_{\mu}$ is closed-valued) and $D_{\mu}$ is $\mathcal{H}$-graph-measurable.
\fdem \smallskip

%
%

It is possible to incorporate measurability in the definition of the essential supremum (see \cite[Section 5.3.1]{KS}  for the definition and the proof of existence of the classical essential supremum). This has been done by  \cite{BCJ} for a single real-valued random variable and by \cite{KL} for a family of vector-valued random variables and with respect to a random partial order (see \cite[Definition 3.1 and Lemma 3.9]{KL}). Proposition \ref{Essup} is given and proved for sake of completeness and for pedagogical purpose. The authors thanks T. Jeulin who suggested this (elegant) proof.
\begin{prop}\label{Essup} Let $\cH\subseteq \cF$ be two $\sigma$-algebras on a probability space. Let $\Gamma=(\gamma_i)_{i\in I}$ be a family of real-valued $\cF$-measurable random variables. There exists a unique $\cH$-measurable random variable $\gamma_{\cH}\in L^0(\R \cup\{\infty\},\cH)$ denoted $\esssup \Gamma$ which satisfies the following properties:
\begin{enumerate}
\item For every $i\in I$, $\gamma_{\cH}\ge \gamma_i$ a.s.
\item If $\zeta \in  L^0(\R \cup\{\infty\},\cH)$ satisfies  $\zeta\ge \gamma_i$ a.s. $\forall i\in I$, then $\zeta\ge \gamma_{\cH}$ a.s.
\end{enumerate}
\end{prop}
The conditional essential infimum $\essinf \Gamma$ is defined symmetrically. \smallskip \\
{\sl Proof.}
Considering the homeomorphism $\arctan$ we can restrict our-self to $\gamma_i$ taking values in $[0,1]$. We denote by $P_{\gamma_i|\cH}$ a regular version of the conditional law of $\gamma_i$ knowing $\cH$. Let $\zeta \in  L^0(\R \cup\{\infty\},\cH)$ such that $\zeta\ge \gamma_i$ a.s. $\forall i\in I$.
This is equivalent to $P_{\gamma_i|\cH}(]-\infty,x])|_{x=\zeta }=1 \mbox{ a.s. }$ and $\supp[\cH]\gamma_i \subset ]-\infty,\zeta ]$ a.s. follows from Definition \ref{DefD}.
Let
\begin{eqnarray}
\label{eqthierry}
\Lambda_{\gamma_i|\cH}=\sup\{x \in [0,1], \, x \in \supp[\cH]\gamma_i\}.
\end{eqnarray}
Then $\Lambda_{\gamma_i|\cH} \leq \zeta $ a.s.
and it  is easy to see that
$\Lambda_{\gamma_i|\cH}$ is $\cH$-measurable. So taking the classical essential supremum, we get that
$\mathrm{ess\,sup}_i \Lambda_{\gamma_i|\cH} \leq \zeta $ a.s. and that $\mathrm{ess\,sup}_i \Lambda_{\gamma_i|\cH}$ is $\cH$-measurable. We conclude that
$\gamma_{\cH}=\mathrm{ess\,sup}_i \Lambda_{\gamma_i|\cH}$ a.s. since for every $i\in I$,
$P(\gamma_i\in  \supp[\cH]\gamma_i|\cH)=1$
(see Remark \ref{remsupp}).
\fdem
\begin{rem}
  \label{essCharact}
{\rm Let $Q$ be an absolutely continuous probability measure with respect to $P$. Let $Z=dQ/dP$ and  $\E_Q$ be the
expectation under $Q$.
As for every $i\in I$, $\esssup \Gamma \ge \gamma_i$ a.s. and $\esssup \Gamma$ is $\cH$-measurable,
\begin{eqnarray}
\label{eqessupespcondi}
\esssup \Gamma \geq \frac{ \E(Z\gamma_i|\cH)}{\E(Z|\cH)}=\E_Q(\gamma_i|\cH).
\end{eqnarray}
}
\end{rem}

Inspired by Theorem 2.8 in \cite{BCJ}, we may easily show the following tower law property.
\begin{lemm}\label{towerprop} Let $\cH_1\subseteq \cH_2\subseteq \cF$ be  $\sigma$-algebras and let $\Gamma=(\gamma_i)_{i\in I}$ be a family of real-valued $\cF$-measurable random variables. Then,
$$\esssup[\cH_1]\left(\esssup[\cH_2] \Gamma\right)=\esssup[\cH_1]\Gamma.$$
\end{lemm}

\begin{lemm}
\label{lemsuppess} Assume that $d=1$ and consider $X\in L^0(\R_+,\cF)$. Then, we have a.s.  that
\begin{eqnarray}
\nonumber
\essinf X&=&\inf {\rm supp}_{\cH}X,\quad
\esssup X=\sup {\rm supp}_{\cH}X,\\
\nonumber
\essinf X&\in& {\rm supp}_{\cH}X,\quad  {\rm on\,\, the\,\, set\,\,} \{\essinf X>-\infty\},\\
\nonumber
\esssup X&\in& {\rm supp}_{\cH}X,\quad  {\rm on\,\, the\,\, set\,\,} \{\esssup X<\infty\},\\
\label{rein}
{\rm conv}{\rm supp}_{\cH}X & = & [\essinf X,\esssup X]\cap \R,
\end{eqnarray}
where ${\rm conv}{\rm supp}_{\cH}X$ is the convex envelop of ${\rm supp}_{\cH}X$, i.e. the smallest convex set that contains ${\rm supp}_{\cH}X$.
\end{lemm}
{\sl Proof.} The two first statements follow from
the construction of $\esssup X$ in Proposition \ref{Essup} (see \eqref{eqthierry}). Suppose that $\essinf X\notin {\rm supp}_{\cH}X$ on some non-null measure subset $\Lambda\in \cH$ of $\{\essinf X>-\infty\}$. As $\supp[\cH]X$ is $\cH$-measurable and closed-valued,  by a measurable selection argument, we deduce the existence of $r\in L^0(\R_+,\cH)$ such that $r>0$ and $(\essinf X-r, \essinf X+r)\subseteq \R\setminus {\rm supp}_{\cH}X$ on $\Lambda$. As $X\in {\rm supp}_{\cH}X$ a.s.  (see Remark \ref{remsupp}) and $X \geq \essinf X$ a.s., we deduce that $X\ge \essinf X+r$ on $\Lambda$, which contradicts the definition of  $\essinf X$. The next statement is similarly shown and the last one follows directly. \fdem

The following proposition is one of the main ingredient of the paper.
It extends the fact that $\esssup[\cH] X=\sup_{x\in \supp[\cH]X} x \mbox{ a.s.}$ (see \eqref{eqthierry}) and allows to compute a conditional essential supremum as a classical supremum but on a random set.
\begin{prop}\label{lemma-essup-h(X)} Let $X \in L^0(\R^d, \cF)$  such that
${\rm dom \; } \supp[\cH]X =\Omega$
and let $h:$ $\Omega \times \R^d \to \R$
be a $\cH \otimes \cB(\R^d)$-measurable function which is  l.s.c.  in $x$.
Then,
\begin{eqnarray}
\label{belequa}
\esssup[\cH] h(X)=\sup_{x\in \supp[\cH]X} h(x) \quad a.s.
\end{eqnarray}
\end{prop}
The proposition has the  following easy extension. The proof is postponed to the appendix.
\begin{coro}\label{lemma-essup-h(set)} Let $\cX \subset L^0(\R^d, \cF)$  such that
${\rm dom \; }\supp[\cH]X =\Omega$ for all $X\in \cX$ and $\cup_{X \in \cX}\supp[\cH]X$ is a $\cH$-measurable and closed-valued random set.
Let $h:$ $\Omega \times \R^d \to \R$
be a $\cH \otimes \cB(\R^d)$-measurable function which is  l.s.c.  in $x$.
Then,
\begin{eqnarray}
\label{belequabis}
\esssup[\cH] \{h(X), \, X \in \cX \}=\sup_{x\in \cup_{X \in \cX}  \supp[\cH]X} h(x) \;\;{\rm a.s.}
\end{eqnarray}
\end{coro}
Note that, if $\cX$ is countable, $\cup_{X \in \cX}\supp[\cH]X$ is clearly $\cH$-measurable. If $\cX= L^0(\R^d, \cF)$, then $\cup_{X \in \cX}\supp[\cH]X=\R^d$, which is again
 $\cH$-measurable and also closed-valued. \\

The proof of Proposition \ref{lemma-essup-h(X)} is based on the  two following useful lemmata.
\begin{lemm}\label{mes} Let $\cK:$ $\Omega \twoheadrightarrow \mathbb{R}^{d}$ be a $\cH$-measurable and closed-valued random set such that ${\rm dom \; }\cK=\O$
and  let $h:$ $\Omega \times \R^d \to \R$
be  l.s.c.  in $x$.  Then,
\begin{eqnarray}
\label{belequa2}\sup_{x\in \cK} h(x)=\sup_{n \in \mathbf{N}} h(\eta_n),
\end{eqnarray}
where $(\eta_n)_{n \in \mathbf{N}}$ is a Castaing representation of $\cK$.
\end{lemm}
{\sl Proof.} Let $\o  \in \O$.
As $(\eta_n(\o))_{n \in \mathbf{N}}\subset \cK(\o)$, $h(\o,\eta_n(\o)) \leq \sup_{x\in \cK(\o)} h(\o,x)$ and thus $\sup_n h(\eta_n) \leq \sup_{x\in \cK} h(x)$.
Let $x \in\cK(\o)={\rm cl}\{\eta_n(\o), \,n\in \mathbf{N}\}$, by lower semicontinuity of $h$
$$h(\o,x) \leq \liminf_n h(\o,\eta_n(\o))\le \sup_n h(\o,\eta_n(\o)).$$
We conclude that  $\sup_{x\in \cK} h(x) \leq \sup_n h(\eta_n)$ and \eqref{belequa2} is proved.
\fdem

\begin{lemm}\label{mesmes} Let $\cK:$ $\Omega \twoheadrightarrow \mathbb{R}^{d}$ be a $\cH$-measurable and closed-valued random set such that ${\rm dom \; }\cK
=\O$
and let $h:$ $\Omega \times \R^k \times \R^d \to \R$ be a $\cH \otimes \cB(\R^k)\otimes \cB(\R^d)$-measurable function such that
  $h(\o,x,\cdot)$  is  l.s.c.  for all
$(\o,x) \in \Omega \times \R^k $. Then
 $(\o,x) \in \Omega \times \R^k  \mapsto s(\o,x)=\sup_{z\in \cK(\o)} h(\o,x,z)$ is $\cH \otimes \cB(\R^k)$-measurable.
\end{lemm}
{\sl Proof.}
Lemma \ref{mes}  implies that
$
s(\o,x)=\sup_n h(\o,x,\eta_n(\o)),
$
where $(\eta_n)_{n \in \mathbf{N}}$ is a Castaing representation of $\cK$.
It implies that  for any fixed  $c \in \R$
\begin{eqnarray*}
\{(\omega,x) \in \O \times \mathbb{R}^{d}, \, s(\o,x)\le c \}
&=&\bigcap_n\{(\omega,x)\in \O \times \mathbb{R}^{d}, \, h(\o,x,\eta_n(\omega))\le c \}.
\end{eqnarray*}
As $h$ is $\cH \otimes \cB(\R^k)\otimes \cB(\R^d)$-measurable and  $\eta_n$ is $\cH$-measurable, $(\omega,x) \mapsto h(\o,x,\eta_n(\omega))$ is $\cH \otimes \cB(\R^k)$-measurable
and so is $s$.
\fdem\smallskip

{\sl Proof of Proposition \ref{lemma-essup-h(X)}.}
As $P(X \in  \supp[\cH]X|\cH)=1$ (see Remark \ref{remsupp}) we have that
$\sup_{x\in \supp[\cH]X} h(x) \geq h(X)$ a.s. and the  definition of
$\esssup[\cH] h(X)$ implies that  $\sup_{x\in \supp[\cH]X} h(x) \geq \esssup[\cH] h(X)$ a.s. since $\sup_{x\in \supp[\cH]X} h(x)$ is $\cH$-measurable by  Lemmata \ref{Dmeasurability} and \ref{mesmes}.

Let $(\gamma_n)_{n \in \mathbf{N}}$ be a   Castaing representation of $\supp[\cH]X.$ Lemma \ref{mes} implies that
$\sup_{x\in \supp[\cH]X} h(x)=\sup_n h(\gamma_n).$
Fix some rational number $\e>0$ and set $Z_{\e}= 1_{B(\gamma_n,\e)}(X)$, where $B(\gamma_n,\e)$ is the closed ball of center $\gamma_n$ and radius $\e$.
Note that  $E(Z_{\e}|\cH)=P(X\in B(\gamma_n,\e)|\cH)>0$. Indeed if it does not hold true $P(X\in \R^d \setminus B(\gamma_n,\e)|\cH)=1$ on some $H \in \cH$ such that $P(H)>0$ and
by definition \ref{DefD}, $\supp[\cH]X \subset \R^d \setminus B(\gamma_n,\e)$ on $H$, which contradicts $\gamma_n\in \supp[\cH]X$. By definition of the essential supremum, we have that $\esssup[\cH] h(X)\geq h(X)$ a.s. and that $\esssup[\cH] h(X)$ is $\cH$-measurable. This implies for all fixed $\o \in \O_{\e}$, where $\O_{\e}$ is of full measure,  that
\begin{eqnarray*}
\esssup[\cH] h(X)(\o)& \geq &  \frac{\E(Z_{\e}h(X)|\cH)}{\E(Z_{\e}|\cH)}(\o) =   \frac{\int 1_{B(\gamma_n(\o),\e)}(x)h(\o,x)  P_{X|\cH}(dx;\o)}{\E(Z_{\e}|\cH)(\o)} \\
& \geq   &   \frac{\int \left(\inf_{y \in B(\gamma_n(\o),\e)}h(\o,y)\right)1_{B(\gamma_n(\o),\e)}(x) P_{X|\cH}(dx;\o)}{\E(Z_{\e}|\cH)(\o)}\\
 & \geq & \inf_{y \in B(\gamma_n(\o),\e)}h(\o,y).
\end{eqnarray*}
As  $h$ is l.s.c. (recall \cite[Definition 1.5, equation 1(2)]{rw}), we have that
$$\lim_{\e \to 0} \inf_{y \in B(\gamma_n,\e)} h(y)=\liminf_{x \to \gamma_n}h(x)=h(\gamma_n).$$
So on the full measure set $\cap_{\e \in \mathbb{Q}, \,\e>0} \O_e$,
$\esssup[\cH] h(X)\geq h(\gamma_n)$. Taking the supremum over all $n$, we get that
$$\esssup[\cH] h(X) \geq \sup_n h(\gamma_n)=\sup_{x\in \supp[\cH]X} h(x) \geq \esssup[\cH] h(X) \mbox{ a.s.}$$\fdem\\

\subsection{Fenchel-Legendre conjugate and bi-conjugate to express super-replication prices and cost}
\label{secgreve}
We are now in position to perform the points 1 to 4 of the program announced in the beginning of the section.
\begin{prop}
\label{lempart1}
\begin{eqnarray}
\label{eqpg}
\cP(g)=\left\{\esssup[\cH]\left(g(Y)-\theta Y \right) + \theta y,~\theta\in L^0(\R^d,\cH)\right\}+L^0(\R_+,\cH).
\end{eqnarray}
Suppose that $g$ is a $\cH$-normal integrand.  Then, for $\theta \in L^0(\R^d,\cH)$, we get that
\begin{eqnarray}
\label{eqf*}
\esssup[\cH]\left(g(Y)-\theta Y \right)=\sup_{z\in {\rm supp}_{\cH}Y}\left(g(z)-\theta z \right)=f^*(-\theta ) \; \; \; {\rm a.s.}
 \end{eqnarray}
where $f^*$ is the Fenchel-Legendre conjugate of $f$ i.e.
\begin{eqnarray}
\nonumber
f^*(\o,x) & = & \sup_{z\in \R^d}\left(xz -f(\o,z)\right)\\
\label{deff}
f(\o,z)& = &-g(\o,z)+\delta_{{\rm supp}_{\cH}Y}(\o,z),
\end{eqnarray}
where $\delta_C(\o,z)=0$ if $z\in C(\o)$ and $+\infty$ else.
Moreover, we have that
 \begin{eqnarray}
 \label{eqbiconj}
p(g)& =& -f^{**}(y) \; \; \; {\rm a.s.}
\end{eqnarray}
where $f^{**}$ is the Fenchel-Legendre biconjugate of $f$ i.e.
$$f^{**}(\o,x)=\sup_{z\in \R^d}\left(xz -f^*(\o,z)\right).$$
 \end{prop}
Notice that the infimum super-hedging cost is not a priori a price, i.e. an element of $\cP(g)$, as the later may be an open interval.\\
\begin{rem}
\label{teemu}
Fenchel-Legendre duality have already been used many times in financial mathematics. In particular, Pennanen obtains a dual representation of the super-replication price thanks to deflators (see \cite[Exemple 4.2]{PenMOR} and \cite[Theorem 10 and Corollary 15]{PenMF}).  The proof of \cite[Theorem 10]{PenMF} is also based on the convex biconjugate theorem but the result is shown under the assumption that the set $\cR$
of claims that can be super-replicate from $0$ (see \eqref{argenttropcher})
is closed, which holds true under the no-arbitrage condition.  In \cite{PenPerkMP}, the existence and the absence of duality gap in a general stochastic optimization problem is proved through dynamic programming and under a condition (that does no rely on inf-compactness) of linearity on sets constructed with recession functions. This condition in classical mathematical finance problems is equivalent to the no-arbitrage condition (see \cite[Exemple 1]{PenPerkMP}). Our approach is different as we do not postulate any assumption on the market and we deduce from the biconjugate representation the condition that should be satisfied by the market. In particular, the goal is not to obtain a dual representation thanks to deflator or martingale measures.
\end{rem}
{\sl Proof.}
As $x\in \cP(g)$ if and only if there exists $\theta\in L^0(\R^d,\cH)$ such that   $x-\theta y \ge g(Y)-\theta Y \, {\rm a.s.}$, we get by definition of the conditional essential supremum (see Proposition \ref{Essup}) that \eqref{eqpg} holds true.
Then \eqref{eqf*}  follows from Proposition \ref{lemma-essup-h(X)}. Lemma \ref{Dmeasurability} will be in force.
First, it implies that  $\delta_{{\rm supp}_{\cH}Y}$  is $\cH\otimes \mathcal{B}(\R^d)$-measurable and  l.s.c.  As ${\rm dom \,}f={\rm supp}_{\cH}Y$ is non-empty (see Remark \ref{remsupp}) $f^*(\o,\cdot)$ is convex and l.s.c. as the supremum of affine functions.  Hence  $x \mapsto f^*(\o,-x)$ is also  l.s.c. and convex. Moreover, using Lemma \ref{mesmes}, $f^*(\o,x)=\sup_{z \in {\rm supp}_{\cH}Y(\o)}\left( xz + g(\o,z)\right)$ is $\cH\otimes \mathcal{B}(\R^d)$-measurable. We obtain that a.s.
 \begin{eqnarray*}
p(g)& =&\essinf[\cH]\{f^*(-\theta)+ \theta y,~\theta\in L^0(\R^d,\cH)\}\\
 & = & -\esssup[\cH]\{\theta y - f^*(\theta),~\theta\in L^0(\R^d,\cH)\}\\
 & = & -\sup_{z\in \R^d }\left(z y-f^*(z)\right)=-f^{**}(y).
\end{eqnarray*}
The first equality is a direct consequence of \eqref{eqpg}, the second one is trivial. We prove  the third one. First,  remark  that $\esssup[\cH]\{\theta y - f^*(\theta),~\theta\in L^0(\R^d,\cH)\}$ coincides with $\esssup[\cH]\{\theta y - f^*(\theta),~\theta\in L^0(\R^d,\cH)\cap {\rm dom \;}f^*\}.$
Moreover, as $f^*$ is  $\cH \otimes \cB(\R^d)$-measurable,
$${\rm graph\,}{\rm dom \;}f^*=\{(\omega,x)\in \Omega\times \R^d, \; f^*(\omega,x)<\infty\} \in \cH \otimes \cB(\R^d)$$  and
${\rm dom \;}f^* $ is $\cH$-measurable (see  \cite[Theorem  14.8]{rw}).
Since $(\o,z) \mapsto z y(\o)-f^*(\o,z)$ is a $\cH \otimes \cB(\R^d)$-measurable function and $f^*(\o,\cdot)$ is convex and thus  u.s.c. on ${\rm dom \,}f^*(\o)$, we may apply Corollary \ref{lemma-essup-h(set)} and we obtain that a.s.
\begin{eqnarray*}
\esssup[\cH]\{\theta y - f^*(\theta),~\theta\in L^0(\R^d,\cH)\cap {\rm dom \; }f^*\} & = & \sup_{z\in{\rm dom}(f^*) }\left(z y-f^*(z)\right)\\
 & = & \sup_{z\in \R^d }\left(z y-f^*(z)\right).
 \end{eqnarray*}
\fdem \smallskip \\ 
We now introduce the notations needed to perform the point 5 of our program. Let $h:\, \R^d \to \R,$ ${\rm conv\,} h$ is the convex envelop of $h$ i.e. the greatest convex function dominated by $h:$
${\rm conv\,} h(x)=\sup\{u(x), \; u \mbox{ convex and } u \leq h \}.$
The concave envelop is defined symmetrically and denoted by ${\rm conc \;} h$. We also define the (lower) closure $\underline{h}$ of $h$ as the greatest l.s.c.  function which is dominated by $h$ i.e.
$\underline{h}(x)=\liminf_{y \to x} h (y)$.  The upper closure is defined symmetrically.
It is easy to see that
$$\underline{{\rm conv }}\,h(y)=\sup{\{\a y + \b, \, \a\in \R^d,\,\b \in \R,\; h(x)\geq \a x + \b,\, \forall x\in\R^d\}}.$$
It is well-known (see for example \cite[Theorem 11.1]{rw}) that
\begin{eqnarray*}
h^*= ({\rm conv}  \,h)^*=(\underline{h})^*=(\underline{{\rm conv}  \,h})^*.
\end{eqnarray*}
Moreover, if ${\rm conv}  \,h$ is proper,  $h^{**}$ is also proper, convex and l.s.c. and
\begin{eqnarray}
\label{eqpasiconcon}
h^{**}= \underline{{\rm conv}  \,h}.
\end{eqnarray}
We are now on position to obtain the representation of the infimum super-hedging cost.
\begin{prop}
\label{fench} Suppose that $g$ is a $\cH$-normal integrand and that there exists some concave function $\varphi$  such that $g\le \varphi$ on ${\rm supp}_{\cH}Y$ \footnote{This is equivalent to assume that there exists $\a, \b \in \R$, such that $g(x)\le \a x + \b $ for all $x \in {\rm supp}_{\cH}Y$.} and
$\varphi < \infty$ on ${\rm conv}{\rm supp}_{\cH}Y$.
Then, a.s.
\begin{eqnarray}
\label{prixavecAIP}
p(g) & = &  \overline{{\rm conc}}(g,{\rm supp}_{\cH}Y)(y)-\delta_{ {\rm conv}{\rm supp}_{\cH}Y}(y)\\
\nonumber
 & = & \inf{\{\a x + \b, \, \a \in \R^d,\,\b \in \R,\; \a z + \b \geq g(z),\, \forall z  \in  {\rm supp}_{\cH}Y\}}\\
 & &  -\delta_{ {\rm conv}{\rm supp}_{\cH}Y}(y),
\end{eqnarray}
where the relative concave envelop of $g$ with respect to ${\rm supp}_{\cH}Y$ is given by
$$
{\rm conc}(g, {\rm supp}_{\cH}Y)(x)=
\inf\{ v(x),~ v\,{\rm\, is\, concave\, and\, }v(z)\ge g(z),\,\forall z\in {\rm supp}_{\cH}Y \}.$$
\end{prop}
Note that  \cite{CGT} and \cite{BN14} have represented the super-hedging price as
a concave envelop  but this was done under the no-arbitrage condition using the dual representation of the super-replication price  through martingale measures. \smallskip  \\
{\sl Proof.}
We want to use \eqref{eqpasiconcon} in order to compute $p(g)$.
The convex envelop of $f$ can be written as follows  (see \cite[Proposition  2.31]{rw}):
\begin{small}
\begin{eqnarray*} {\rm conv \,}f(x)   =&  \inf\left\{\sum_{i=1}^n \l_i f(x_i), \,  n \geq 1,\,  (\l_i)_{i\in \{1,\ldots,n\}} \in \R_+^n, \,  (x_i)_{i\in \{1,\ldots,n\}} \in \R^{d \times n}, \right. \\
 & \left. x=\sum_{i=1}^n \l_i  x_i, \, \sum_{i=1}^n \l_i=1\right\}.
 \end{eqnarray*}
\end{small}
Let $x=\sum_{i=1}^n \l_i x_i$ for some $ n \geq 1,\,  (\l_i)_{i\in \{1,\ldots,n\}} \in \R_+^n$ such that $\sum_{i=1}^n \l_i=1$ and  $(x_i)_{_{i\in \{1,\ldots,n\}}} \in \R^{d \times n}$.
Assume that $x \notin {\rm conv}{\rm supp}_{\cH}Y$. Then (see \cite[Proposition  2.27, Theorem 2.29]{rw}),
there exists at least one $x_i \notin {\rm supp}_{\cH}Y$ and $f(x_i)=+\infty$ and also ${\rm conv \,}f(x)=+\infty$.
If  $x \in {\rm conv}{\rm supp}_{\cH}Y$, by definition ${\rm conv \,}f(x)=-{\rm conc}(g, {\rm supp}_{\cH}Y)(x)$. \\
As  ${\rm conv}{\rm supp}_{\cH}Y$ is non-empty (see Remark \ref{remsupp}), ${\rm conv \,}f$ is  proper if and only if ${{\rm conc}}(g, {\rm supp}_{\cH}Y)(x)<+\infty$  for all $ x\in {\rm conv}{\rm supp}_{\cH}Y$ and this holds true since
$${\rm conc}(g, {\rm supp}_{\cH}Y)\le \varphi <\infty \mbox{ on }{\rm conv}{\rm supp}_{\cH}Y.$$
As for  all $x\in {\rm conv}{\rm supp}_{\cH}Y$,
${{\rm conc}}(g, {\rm supp}_{\cH}Y)(x)\geq g(x)>-\infty, $  we get that   ${\rm conc}(g, {\rm supp}_{\cH}Y)(x) \in \R$
and one may write that \bean
{\rm conv \,}f &=&-{\rm conc}(g, {\rm supp}_{\cH}Y) +\delta_{{\rm conv}{\rm supp}_{\cH}Y} \; \; \; {\rm a.s.}
\eean
and  using Proposition \ref{lempart1} and \eqref{eqpasiconcon}
\begin{eqnarray*}
p(g)    &=&-f^{**}(y)=-\underline{{\rm conv}} \,f(y)  \\
& =& {\overline{\rm conc}}(g, {\rm supp}_{\cH}Y)(y) - \delta_{{\rm conv}{\rm supp}_{\cH}Y}(y) \; \; \; {\rm a.s.}
\end{eqnarray*}


 \fdem \\

\subsection{The AIP condition}
\label{secAIP}
Proposition \ref{fench} shows that if, $ y\notin {\rm conv}{\rm supp}_{\cH}Y$, the infimum super-hedging price of a European claim $p(g)$ equals $-\infty$. This leads to the natural  notion of absence of immediate profit that we present now.  It is important to note that this notion is endogenous to the problem of super-replication contrary to the NA condition.
Let $\cR$ be the set of all $\cF$-measurable claims that can be super-replicate from $0$.
\begin{eqnarray}
\label{argenttropcher}
\cR=\left\{ \theta(Y-y)-\epsilon^+,~\theta \in L^0(\R^d,\cH),\;\epsilon^+\in L^0(\R_+,\cF)\right\}.
\end{eqnarray}
Then,
\begin{eqnarray*}
\cP(0) &= & \{x\in  L^0(\R,\cH),  \exists\, \theta\in L^0(\R^d,\cH),\;   x+\theta (Y -y)\ge 0 \, {\rm a.s.}\} \\
&= & (-\cR) \cap L^0(\R,\cH).
\end{eqnarray*}
Note that $0\in  \cP(0),$ so $p(0) \leq 0.$
We say that there is an immediate profit when $P(p(0)<0)>0$ i.e. if it is possible to super-replicate the contingent claim $0$ at a negative super-hedging price.
 \begin{defi}
 \label{defipone}
 There is an immediate profit (IP)  if
 $P(p(0)<0)>0$.  On the contrary case  if  $p(0)=0$ a.s. we say that the Absence of Immediate Profit (AIP) condition holds.
 \end{defi}

We know propose several characterisations of the AIP condition. We will discuss in Lemma \ref{remoa} and Remark \ref{remoa1}, the link with the classical no-arbitrage condition and show that AIP is indeed very week.

\begin{prop}
\label{NGDone}
AIP holds  if and only if one of the following condition holds true.
\begin{enumerate}
\item  $y \in {\rm conv}{\rm supp}_{\cH}Y$ a.s.  or $0 \in {\rm conv}{\rm supp}_{\cH}(Y-y)$ a.s.
\item $\sigma_{{\rm supp}_{\cH}(Y-y)} \geq 0$ a.s. where $\sigma_{D}(z)=\sup_{x \in D} (-xz)$ is the support function of $-D$
\item $\cP(0) \cap L^0(\R_-,\cH)=\{0\}$  or $\cR \cap L^0(\R_+,\cH)=\{0\}.$
\end{enumerate}
\begin{rem}
\label{foi}
In the case $d=1$,  \eqref{rein} implies that the previous conditions are equivalent to
$y \in  [\essinf[\cH] Y, \esssup[\cH]  Y] \cap \R\, {\rm a.s.}$
\end{rem}
\begin{ex}
The AIP  condition is very easy to check in practice. For $d=1$, let $Z=Y/y$. To check AIP,  compute either ${\rm supp}_{\cH}Z$ or $\essinf[\cH] Z$ and  $\esssup[\cH]  Z$ and compare with $1$.  For example, let  $Z=e^{(\mu -\frac{\s^2}2)+ \sigma(B_{t+1}-B_t)}$ where $(B_t)_{t \in \R_+}$ is a Brownian motion and $\cH=\sigma(\{B_u,\, 0\leq u \leq t\})$ and
$\cF=\sigma(\{B_u,\, 0\leq u \leq t+1\})$. Then ${\rm supp}_{\cH}Z=[0,\infty)$ and  AIP holds true. We propose in Example \ref{exto} other situation  where AIP is easily verified.
\end{ex}
\end{prop}
{\sl Proof.}
The assumptions of Proposition \ref{fench} are satisfied for $g=0$
and we get that
$p(0)=-\delta_{{\rm conv}{\rm supp}_{\cH}Y}(y)$ a.s. Hence, AIP holds true if and only if $ y\in  {\rm conv}{\rm supp}_{\cH}Y$ a.s. or equivalently $0 \in {\rm conv}{\rm supp}_{\cH}(Y-y)$ a.s.  and AIP is equivalent to 1.\\
Using Proposition \ref {lempart1}, we get that
\begin{eqnarray*}
\cP(0) & =& \left\{\esssup[\cH]\left(-\theta (Y-y) \right),~\theta\in L^0(\R^d,\cH)\right\}+L^0(\R_+,\cH).
\end{eqnarray*}
Proposition \ref{lemma-essup-h(X)} implies that for $\theta\in L^0(\R^d,\cH)$
$$\esssup[\cH]\left(-\theta (Y-y) \right)=\sup_{x \in  {\rm supp}_{\cH}(Y-y)}\left(-\theta x \right)=\sigma_{{\rm supp}_{\cH}(Y-y)}(\theta).$$
So, $\cP(0) \cap L^0(\R_-,\cH)=\{0\}$  if and only if $\sigma_{{\rm supp}_{\cH}(Y-y)} \geq 0$ a.s. and $2.$ and $3.$ are equivalent. To achieve the proof, it remains to prove that  $\sigma_{{\rm supp}_{\cH}(Y-y)} \geq 0$ a.s. is equivalent to
$0 \in {\rm conv}{\rm supp}_{\cH}(Y-y)$ a.s.  First remark that
$$\sigma_{{\rm supp}_{\cH}(Y-y)}= \sigma_{{\rm conv}{\rm supp}_{\cH}(Y-y)}.$$
So, it remains to prove that for any  closed convex set $D \in \R^d$,  $\sigma_D \geq 0$ if and only if $0 \in D$.  If  $0 \in D$ it is clear that
$\sigma_D \geq 0$. Assume that $0 \notin D$. Then by Hahn-Banach theorem there exists some $\b>0$ and some $\theta_0 \in \R^d \setminus\{0 \}$ such that
$-x \theta_0 \leq -\beta$ for all $x \in D$ and $\sigma_D(\theta_0) \leq -\beta<0$ follows.
\fdem\\

\begin{coro}
\label{aipcall} The AIP condition holds true if and only if $p(g) \geq 0$ a.s.  for some  non-negative $\cH$-normal integrand $g$ such that there exists some concave function $\varphi$  verifying that $g\le \varphi < \infty$.
\end{coro}
In particular, the AIP condition holds true if and only if the infimum super-hedging cost of some European call option is non-negative. Note that under AIP the price of some non-zero call option may be zero (see Example \ref{calc} below).\\
{\sl Proof.}
Assume that AIP condition holds true. Then, from Definition \ref{defipone}, we get that $p(0)=0$ a.s. As $g \geq 0$, it is clear that $p(g) \geq p(0)=0$ a.s.
Conversely, assume that there exists some IP. Proposition  \ref{fench} implies that
$$p(g)  =\overline{{\rm conc}}(g,{\rm supp}_{\cH}Y)(y)-\delta_{ {\rm conv}{\rm supp}_{\cH}Y}(y).$$
IP and Proposition \ref{NGDone} lead to  $P(y \in {\rm conv}{\rm supp}_{\cH}Y)<1$ and, since
$$\overline{{\rm conc}}(g,{\rm supp}_{\cH}Y)(y) \leq \varphi<\infty,$$ $P(p(g)=-\infty)>0.$ The converse is proved.
\fdem

We now compare  the AIP condition with the classical No Arbitrage  NA one, whose definition is recalled below.
 \begin{defi}
 \label{defiNA}
 The No Arbitrage  NA condition holds true if for $\theta \in L^0(\R^d,\cH)$, $\theta(Y- y) \geq 0$ a.s. implies that $\theta(Y-y) = 0$ a.s. or equivalently  if $ \cR\cap L^0(\R_+,\cF)=\{0\}.$
 \end{defi}
 \begin{lemm}
\label{remoa}
The AIP condition is strictly weaker than the  NA one.
\end{lemm}
\begin{rem}
\label{remoa1}
The AIP condition is tailor-made for pricing issues. It allows to give a super-hedging price even in case of arbitrage opportunity (see example \ref{calc} below).
Note that an IP is a very strong strategy. Assume that $\cH$ is trivial, then an IP corresponds to some $\theta \in \R^d$ such that $\theta(Y-y)$ is determinist and strictly positive. So excluding IP and not NA may be not enough to get existence in the problem of  maximization of expected utility. 
We compare IP with other notions of arbitrage as introduced by Ingersoll (see \cite{Ing}) in a one step setting with a finite set of states of the world. Arbitrage opportunity of the first type is the classical arbitrage.  An arbitrage opportunity $\theta$ of the second type
is limited liability investments with a current negative commitment. As we assume the existence of a riskless asset, it means that $\theta(Y-y)$ is not determinist but always greater that some strictly positive deterministic number. Finally, a riskless arbitrage
opportunity is a nonpositive investment with a constant, positive profit. This notion is equivalent to our notion of IP (recall that there exits a riskless asset) in the context of a trivial initial filtration. If $\cH$ is not trivial anymore, a riskless arbitrage is an IP but the converse is not true anymore. \\
An unbounded profit with bounded risk is some $\theta \in L^0(\R^d,\cH)$ such that $P(\theta(Y-y) \geq 0)=1$ and $P(\theta(Y-y) > 1)>0.$ Let us show that one can have AIP and some unbounded profit with bounded risk. Fix $d=1$ and choose some random variables $Y$ and $y$ such that $\essinf[\cH]  Y=y$ and $P(\Gamma)>0$ where
$\Gamma=\{\esssup[\cH] Y > y+1\}.$ Here, AIP holds true (recall Remark \ref{foi}). Observe that
$(Y- y) \geq \essinf[\cH] Y-y= 0$ a.s. Now if  $Y-y \leq 1$ a.s. then $P({\Gamma})=0$, i.e. a contradiction. So the  constant strategy equal to 1 is an unbounded profit with bounded risk.
\end{rem}
{\sl Proof.}
It is clear from Proposition \ref{NGDone} and Definition \ref{defiNA} that NA implies  AIP. Fix $d=1$ and choose some random variables $Y$ and $y$ such that $\essinf[\cH]  Y=y$ and $P(\Gamma)>0$ where
$\Gamma=\{\esssup[\cH] Y > y\}.$ Here, AIP holds true (recall Remark \ref{foi}). Observe that
$(Y- y) \geq \essinf[\cH] Y-y= 0$ a.s. Now if  $Y-y=0$ a.s. then $P({\Gamma})=0$, i.e. a contradiction. So the  constant strategy equal to 1 is an arbitrage opportunity.
\fdem \\

We propose now a condition for the equivalence between NA and AIP when $d=1$.
\begin{lemm}
\label{NAAIP}
Assume that $d=1$ and that  $P(\essinf[\cH]  Y=y)=P(\esssup[\cH]Y=y)=0$. Then AIP  and NA are equivalent conditions.
\end{lemm}
Lemma \ref{NAAIP} applies if $\essinf[\cH]  Y=0$, $\esssup[\cH]  Y=\infty$ and $y\in (0,\infty)$.\\

\noindent {\sl Proof.}
We have already seen that NA implies AIP. Assume that AIP holds true. Using Remark  \ref{foi},
$y \in  [\essinf[\cH] Y, \esssup[\cH]  Y] \cap \R\, {\rm a.s.}$ 
Let $\theta \in L^0(\R,\cH)$ such that  $\theta(Y-y)\ge 0$. On the set $\{\theta>0\}\in \cH$, we have that $Y\ge y$ hence $\essinf[\cH] Y\ge y \geq \essinf[\cH] Y$. We deduce that $P(\theta>0)=0$. Similarly, we get that $P(\theta<0)=0$ and finally $\theta=0$.
\fdem \\
\begin{ex}
\label{exto}
We now provide an other example where AIP holds true and  is strictly weaker than NA in the case $d=1$. First, notice that, if there exists $Q_1,Q_2\ll \bP$ such that $(y,Y)$ is a $Q_1$-super martingale and a $Q_2$-sub martingale, then  AIP holds true. Indeed, let $Z_i=dQ_i/dP$ for  $i \in  \{1,2\}.$ As $\essinf Y \leq Y \leq \esssup Y$ a.s., $\esssup Y$ and $\essinf Y$ are  $\cH$-measurable, we get that a.s. $\esssup Y \geq \E_{Q_2}(Y|\cH)\geq y$ (see \eqref{eqessupespcondi})  and $\essinf Y \geq \E_{Q_1}(Y|\cH)\leq y$. So   Remark \ref{foi} implies that AIP holds true.\smallskip

Let us consider  $M\in L^0((0,\infty),\cF)$, such that  $\essinf(M)<M<\esssup(M)$ a.s.   We define $Y:=M-\essinf(M)>0$ and $y=\alpha_H\esssup(M)-\essinf(M)>0$ where $\alpha_H\in L^0(\R,\cH)$ is chosen such that $\alpha_H \in (\frac{\essinf(M)}{\esssup(M)},1]$ a.s. Morever, we suppose that $\alpha_H=1$ on a non null set $A_{\cH}\in \cH$ that we arbirarily choose. By construction, AIP holds, since $\essinf(Y)=0<y$ and $\esssup(Y)=\esssup(M)-\essinf(M)\ge y$. Suppose that NA holds, then by the FTAP, there exists $Q\sim P$ such that $E_{Q}(Y|\cH)=y$. This implies that 
$$y=E_{Q}(M|\cH)-\essinf(M)=\alpha_H\esssup(M)-\essinf(M).$$
In particular, we have $E_{Q}(M|\cH)=\esssup(M)$ on $A_{\cH}$. This contradicts the hypothesis $M<\esssup(M)$ a.s. We may also show directly that  $\theta=-1_{A_{\cH}}$ is an arbitrage opportunity. Indeed 
$-1_{A_{\cH}}(Y-y)=1_{A_{\cH}}(\esssup(M)-M)$ which is a.s. stricly positive.
\fdem   \smallskip
%
\end{ex}

We now provide the characterization of the infimum super-hedging cost under the AIP condition.
\begin{coro}
\label{fenchngd} Suppose that AIP holds true. Let $g$ be a $\cH$-normal integrand, such that there exists some concave function $\varphi$  verifying that $g\le \varphi$ on ${\rm supp}_{\cH}Y$ and  $\varphi < \infty$ on ${\rm conv}{\rm supp}_{\cH}Y$.
Then, a.s.
\begin{eqnarray}
p(g)&=&\overline{{\rm conc}}(g, {\rm supp}_{\cH}Y)(y)\\ \nonumber
\label{eqfenchprix}
&=&\inf{\{\a y + \b, \, \a \in \R^d,\,\b \in \R,\;  \a x + \b\geq g(x),\, \forall x\in {\rm supp}_{\cH}Y\}}.
\end{eqnarray}
If $g$ is concave and u.s.c., $p(g)=g(y)$ a.s.
\end{coro}
{\sl Proof.}
The first equalities  are a direct consequence of Proposition \ref{fench}. If $g$ is concave and u.s.c., the result is trivial.\fdem \smallskip \\
We finish the one-period study with the computation of the infimum super-hedging cost of a convex derivative when $d=1$.
In this case, the cost is in fact a super-hedging price and we get the super-hedging strategy explicitly.
\begin{coro}
\label{fenchngdconv} Suppose that AIP holds true and that $d=1$. Let $g:\R \to \R$ be a non-negative convex function with ${\rm dom} \, g=\R$ and $\lim_{x\to \infty} x^{-1}g(x)=M \in [0,\infty)$, then a.s.
\begin{eqnarray}\label{PricingFormula}
p(g) & = &  \theta^* y + \beta^*=g(\essinf[\cH] Y ) + \theta^*\left( y-\essinf[\cH]Y \right),\\
 \label{theta}
\theta^* & = & \frac{g(\esssup[\cH] Y )-g(\essinf[\cH]Y)}{\esssup[\cH] Y-\essinf[\cH] Y},
\end{eqnarray}
where we use the conventions $\theta^*=\frac{0}{0}=0$ in the case $\esssup[\cH] Y=\essinf[\cH] Y$ a.s. and $\theta^*=\frac{g(\infty)}{\infty}=M$ if $\essinf[\cH] Y<\esssup[\cH] Y=+\infty$ a.s. Moreover, $p(g)\in \cP(g)$.
\end{coro}
\begin{ex}
\label{calc}
We compute the price of a call option under AIP in the case $d=1$. Let $G=g(Y)=(Y-K)_+$ for some $K\geq 0$. 
\begin{itemize}
\item If $K\geq \esssup[\cH] Y$ then $Y-K\leq \esssup[\cH] Y -K$ and $G=0$. As AIP condition holds true, $p(g)=p(0)=0$. 
\item If $K\leq \essinf[\cH] Y$ then $Y-K\geq \essinf[\cH] Y -K$ and $G=Y-K$. As $g$ is concave and u.s.c., $p(g)=g(y)= y-K$ a.s.  
\item If $\essinf[\cH] Y\leq K \leq \esssup[\cH] Y.$ Then \eqref{theta} and \eqref{PricingFormula} imply that
\begin{eqnarray*}
p(g) & = &  \frac{\esssup[\cH] Y -K }{\esssup[\cH] Y-\essinf[\cH] Y}\left( y-\essinf[\cH]Y \right)
\end{eqnarray*}
on $\{\esssup[\cH] Y\neq \essinf[\cH] Y\}$ and 0 else. 
So $p(g)=0$ if and only if  $y=\essinf[\cH]Y$ or $\esssup[\cH] Y= \essinf[\cH] Y$. A non-negative call option can have a zero price.
\end{itemize}
We finish with an example of computation of a call price under AIP but when there is some arbitrage opportunity. We choose a simple model that will be studied in Section \ref{seexpli}. We assume that $\essinf[\cH]Y=dy$ a.s. and  $\esssup[\cH]Y=u y$ a.s. for two constants $u$ and $d$. From Remark \ref{foi},
AIP is equivalent to $d\in[0,1]$ and $u\geq 1$. If $d=1$ (and $u>1$) or $u=1$ and ($0\leq d<1$),   AIP holds but the NA condition does not hold true.   Suppose that $d=1$ and $u>1$. If $K\in [y,\infty)$, the super-replication price under AIP is $0$ and if $K \leq y$ it is $y-K$. Suppose that $u=1$ and $0\leq d<1$. If $K\in [0,y]$, the super-replication price under AIP is $y-K$ and if $K\geq y$ it is zero.
\end{ex}
{\sl Proof.}
As $g$ is convex, the relative concave envelop of $g$ with respect to ${\rm supp}_{\cH}Y$ is the affine function that coincides with $g$ on the extreme points of the interval ${\rm conv}{\rm supp}_{\cH}Y$ and \eqref{PricingFormula} and \eqref{theta} follow from Remark \ref{foi}. Then using \eqref{eqfenchprix}, we get that
$\theta^* Y + \beta^* \geq g(Y)$ a.s. (recall that $Y \in {\rm supp}_{\cH}Y$)
and this implies by
\eqref{PricingFormula} that
\begin{eqnarray}\label{lavieestbelle}
p(g)+ \theta^* (Y-y) \geq g(Y) \; {\rm a.s.}
\end{eqnarray}
and $p(g)\in \cP(g)$ follows.
\fdem \\

\section{The multi-period framework}
\label{secmulti}
\subsection{Multi-period super-hedging prices}
For every $t \in \{0,\ldots,T\},$ the set $\mathcal{R}_t^T$ of all  claims that can be super-replicated from the zero initial endowment at time $t$ is defined by
\begin{eqnarray}
\label{defR}
\mathcal{R}_t^T:=\left\{\sum_{u=t+1}^T \theta_{u-1}\Delta S_u-\epsilon_T^+,~\theta_{u-1}\in L^0(\R^d,\cF_{u-1}),\;\epsilon_T^+\in L^0(\R_+,\cF_{T})\right\}.\quad
\end{eqnarray}
The set of (multi-period) super-hedging prices and the (multi-period) infimum super-hedging cost  of some contingent claim $g_T\in L^0(\R,\cF_T)$ at time $t$  are given by for all $t \in \{0,\ldots,T\},$ by
\begin{eqnarray}
\nonumber
\cP_{T,T}(g_T)& = & \{g_T\} \mbox{ and }
\p_{T,T}(g_T)  = g_T\\
\label{defPi}
\cP_{t,T}(g_T)   &= &  {\{x_t\in L^0(\R,\cF_{t}),\, \exists R \in \mathcal{R}_t^T,\, x_t+R=g_T  \mbox{ a.s.}\}} \\
\nonumber
\p_{t,T}(g_T) & = &\essinf[\cF_t]\cP_{t,T}(g_T).
\end{eqnarray}
As in the one-period case, it is clear that the infimum super-hedging cost is not necessarily a price in the sense that $\p_{t,T}(g_T)\notin  \cP_{t,T}(g_T)$ when $ \cP_{t,T}(g_T)$ is not closed.

We now define a local version of super-hedging prices. Let $g_{t+1} \in  L^0(\R,\cF_{t+1})$, then the set of one-step super-hedging prices of  $g_{t+1}$ and it associated infimum super-hedging cost are given by
\begin{eqnarray*}
\cP_{t,t+1}(g_{t+1}) & =& \left\{x_t\in  L^0(\R,\cF_{t}),  \exists\, \theta_t\in L^0(\R^d,\cF_{t}),\;   x_t+\theta_t  \Delta S_{t+1} \ge g_{t+1} \, {\rm a.s.} \right\}\\
\pi_{t,t+1}(g_{t+1}) & =& \essinf[\cF_t] \cP_{t,t+1}(g_{t+1}).
\end{eqnarray*}
The following lemma makes the link between local and global super-hedging prices under the assumption that the infimum (global) super-replication cost is a price. It also provides a dynamic programming principle.
\begin{lemm}
\label{lemouf}
Let $g_T \in L^0(\R,\cF_{T})$ and $t \in \{0,\ldots,T\}$. Then
$$\cP_{t,T}(g_T)\subset \cP_{t,t+1}(\pi_{t+1,T}(g_T)) \mbox{ and }
\pi_{t,T}(g_T) \geq \pi_{t,t+1}(\pi_{t+1,T}(g_T)).$$
Moreover, assume that
$\pi_{t+1,T}(g_T) \in \cP_{t+1,T}(g_T)$. Then $$\cP_{t,T}(g_T)=\cP_{t,t+1}(\pi_{t+1,T}(g_T)) \mbox{ and }
\pi_{t,T}(g_T)=\pi_{t,t+1}(\pi_{t+1,T}(g_T)).$$
\end{lemm}
\begin{rem} {\rm We will give in Proposition \ref{propEqWNFL-AIP} conditions under which we have $\pi_{t+1,T}(g_T) \in \cP_{t+1,T}(g_T)$.
Under AIP, if at each step, $\pi_{t+1,T}(g_T) \in \cP_{t+1,T}(g_T)$ and if $\pi_{t+1,T}(g_T)=g_{t+1}(S_{t+1})$ for some ``nice'' $\cF_{t}$-normal integrand $g_{t+1}$, we will get from Corollary \ref{fenchngd} that $\pi_{t,T}(g_T)=\overline{{\rm conc}}(g_{t+1}, {\rm supp}_{\cF_{t}}S_{t+1})(S_{t})$ a.s. We will propose in Section \ref{seexpli} a quite general setting where this holds true.}
\end{rem}
{\sl Proof.}
Let  $\Pi_{T,T}=\{g_T\}$ and for all $t \in \{0,\ldots,T-1\}$
\begin{small}
\begin{eqnarray*}
\Pi_{t,T}(g_T)
 \!\!\!\!&= &\!\!\!\!{\{x_t\in L^0(\R,\cF_{t}),\; \exists \theta_t \in L^0(\R^d,\cF_{t}),\, \exists p_{t+1}\in  \Pi_{t+1,T}(g_T),\,
  x_t+ \theta_t  \Delta S_{t+1} \geq p_{t+1} \mbox{ a.s.}\}}.
\end{eqnarray*}
\end{small}
The set $\Pi_{t,T}(g_T)$ contains  at time $t$ all the super-hedging prices for some price $p_{t+1}\in  \Pi_{t+1,T}(g_T)$ at time $t+1$.
First we prove that for all $t \in \{0,\ldots,T\}$
\begin{eqnarray}
\label{eqcestegal}
\cP_{t,T}(g_T) &= &\Pi_{t,T}(g_T).
\end{eqnarray}
It is clear at time $T$. Let  $t \in \{0,\ldots,T\}$.
Let $x_t \in \cP_{t,T}(g_T)$. Then there exists for all $u\in \{t,\ldots,T-1\}$,  $\theta_{u}\in L^0(\R^d,\cF_{u})$   such that $x_t+\sum_{u=t+1}^{T-1}  \theta_{u-1} \Delta S_u +\theta_{T-1} \Delta S_T \geq g_T$ a.s. So
$$x_t+\sum_{u=t+1}^{T-2}  \theta_{u-1} \Delta S_u +\theta_{T-2} \Delta S_{T-1}=x_t+\sum_{u=t+1}^{T-1}  \theta_{u-1} \Delta S_u  \in \Pi_{T-1,T}(g_T) $$ and
$x_t+\sum_{u=t+1}^{T-2}  \theta_{u-1} \Delta S_u \in \Pi_{T-2,T}(g_T)$ and recursively $x_t \in \Pi_{t,T}(g_T)$. Conversely, let $x_t \in \Pi_{t,T}(g_T)$, then
there exists $\theta_t \in L^0(\R^d,\cF_{t})$ and $p_{t+1}\in  \Pi_{t+1,T}(g_T),$ such that $x_t+ \theta_t  \Delta S_{t+1} \geq p_{t+1} \mbox{ a.s.}$ Then as $p_{t+1}\in  \Pi_{t+1,T}(g_T),$ there exists $\theta_{t+1} \in L^0(\R^d,\cF_{t+1})$ and $p_{t+2}\in  \Pi_{t+2,T}(g_T),$ such that $p_{t+1}+ \theta_{t+1}  \Delta S_{t+2} \geq p_{t+2} \mbox{ a.s.}$ Going forward until $T,$
$p_{T-1}+ \theta_{T-1}  \Delta S_{T} \geq g_T \mbox{ a.s.}$,  we get that $x_t+\sum_{u=t+1}^{T}  \theta_{u-1} \Delta S_u \geq g_T$ a.s. and $x_t \in \cP_{t,T}(g_T)$ follows. This achieve the proof of \eqref{eqcestegal}.

Let $x_t \in \cP_{t,T}(g_T)$, then there exists $\theta_t \in L^0(\R^d,\cF_{t})$ and $p_{t+1}\in  \cP_{t+1,T}(g_T)$ such that (recall \eqref{eqcestegal})
$$x_t+ \theta_t  \Delta S_{t+1} \geq p_{t+1} \geq \essinf[\cF_t]\cP_{t+1,T}(g_T)=\pi_{t+1,T}(g_T) \mbox{ a.s.}$$
and the first statement follows. The second one follows directly from \eqref{eqcestegal} and $\pi_{t+1,T}(g_T) \in \cP_{t+1,T}(g_T)$.
\fdem
\subsection{Multi-period AIP}
We now define the notion of global and local immediate profit at time $t$. The global (resp. local) profits mean that it is possible to super-replicate from a negative cost at  time $t$ the claim $0$ payed at time $T$ (resp. time $t+1$). We will see that they are equivalent.
\begin{defi}
\label{foire}
Fix  $t \in \{0,\ldots,T\}$. A global immediate profit (IP) at time $t$ is a non-null element of $\cP_{t,T}(0) \cap L^0(\R_-,\cF_t)$. We say that
AIP condition holds at time $t$ if there is no global IP at  $t$:
$$\cP_{t,T}(0)\cap L^0(\R_-,\cF_t)=\{0\}.$$
A local immediate profit (LIP) at time $t$ is  a  non-null element of $\cP_{t,t+1}(0)\cap L^0(\R_-,\cF_t)$. We say that
(ALIP) condition holds at time $t$ if there is no local IP at  $t$: $$\cP_{t,t+1}(0)\cap L^0(\R_-,\cF_t)=\{0\}.$$
Finally we say that the AIP (resp. ALIP) condition holds true if the AIP (resp. ALIP) condition holds at time $t$ for all  $t \in \{0,\ldots,T\}$.
\end{defi}
Theorem \ref{thoNip} below proposes several characterization of the (AIP) condition.
\begin{theo}\label{thoNip} AIP holds  if and only if one of the the following assertions holds.\smallskip
\begin{enumerate}
\item ALIP holds true.\smallskip

\item $S_t \in {\rm conv}{\rm supp}_{\cF_t}S_{t+1}\, {\rm a.s.}$  or
$0 \in {\rm conv}{\rm supp}_{\cF_t}(S_{t+1}-S_t)\, {\rm a.s.}$ for all $t\in \{0,\ldots,T-1\}.$\smallskip

\item $\s_{{\rm supp}_{\cF_t}(S_{t+1}-S_t)} \geq 0\, {\rm a.s.}$  for all $t\in \{0,\ldots,T-1\}.$\smallskip

\item $\p_{t,T}(0)=0$ \, {\rm a.s.} for all $t\in \{0,\ldots,T\}.$
\end{enumerate}

\end{theo}
\begin{rem}
In the case $d=1$, the previous conditions are equivalent to
\begin{itemize}
\item $\essinf[\cF_{t}]  S_{t+1}\le S_t \le \esssup[\cF_{t}]  S_{t+1}\, {\rm a.s.}$  for all $t\in \{0,\ldots,T-1\}.$
\item $\essinf[\cF_{t}]  S_{u}\le S_t \le \esssup[\cF_{t}]  S_{u}$ {\rm a.s.} for all $u\in \{u,\ldots,T\}$.
\end{itemize}
The equivalence between 2. and the first (resp. second) item comes from \eqref{rein}  (resp.  Lemma \ref{towerprop}).
\end{rem}
{\sl Proof.}
At time $T$,  $\cP_{T,T}(0)=\{0\}$, thus AIP holds at $T$ and $\pi_{T,T} (0)=0$.
We show by induction that if $0\in \cP_{t+1,T}(0)$ and if AIP  at time $t+1$ holds true then $0\in \cP_{t,T}(0)$ and the following equivalences are true :
\begin{eqnarray*}
\p_{t,T}(0)=0 \, {\rm a.s.} & \Leftrightarrow  & S_t \in {\rm conv}{\rm supp}_{\cF_t}S_{t+1}\, {\rm a.s.} \Leftrightarrow  \s_{{\rm supp}_{\cF_t}(S_{t+1}-S_t)} \geq 0\, {\rm a.s.}\\
&\Leftrightarrow & \mbox{AIP holds at time } t
\Leftrightarrow  \mbox{ALIP holds at time } t.
\end{eqnarray*}
As AIP is equivalent to  AIP  at time $t$ for all  $t \in \{0,\ldots,T\}$, this proves the equivalence between AIP, 1., 2. , 3.  and 4.
Consider $t\in \{0,\ldots,T-1\}$, assume that the induction hypothesis holds true at $t+1$, $0\in \cP_{t+1,T}(0)$ and that AIP holds at time $t+1$.
As $\p_{t+1,T}(0)=0 \in \cP_{t+1,T}(0)$,  Lemma \ref{lemouf} shows that $
\cP_{t,T}(0)  =  \cP_{t,t+1}(0)$  and
$\pi_{t,T}(0)=\pi_{t,t+1}(0)$.
This implies that  AIP at time $t$ is equivalent to ALIP at time $t$ and together with Proposition \ref{NGDone} and Definition \ref{defipone} shows that the induction step holds at time $t$ and that $\pi_{t,t+1}(0)=0 \in \cP_{t,t+1}(0)=\cP_{t,T}(0)$.
\fdem


\subsection{Absence of weak immediate profit}
\label{secomp}
In this section we study a  condition stronger than AIP in the spirit of the No free Lunch condition i.e.
by considering the closure of the set $\mathcal{R}_t^T$. Before,  we  recall the classical multiperiod no-arbitrage NA condition.
 \begin{defi}
 \label{defiNAmulti}
 The no-arbitrage  NA condition holds  if   for all $t \in \{0,\ldots, T\},$
$$\mathcal{R}_t^T \cap L^0(\R_+,\cF_T)=\{0\}.$$
 \end{defi}
It is easy to see that the NA condition can also be formulated as follows~:   $V_T^{0,\theta} \geq 0$ a.s. implies that $V_T^{0,\theta}= 0$ a.s. Recall that the set of all super-hedging  prices for the zero claim at time $t$ is given by $\cP_{t,T}(0)=(-\mathcal{R}_t^T)\cap L^0(\R,\cF_t)$ (see \eqref{defR} and \eqref{defPi}). It follows that (see Definition \ref{foire})
$$\mbox{AIP reads as } \mathcal{R}_t^T \cap L^0(\R_+,\cF_t)=\{0\} \mbox{ for all } t \in \{0,\ldots, T\}.$$
It is clear that the NA condition implies the AIP one and, by the counter-example of Lemma  \ref{remoa}, the equivalence does not hold true:
The AIP condition is strictly weaker than the  NA one.  We now introduce  a weaker form of IP.
 \begin{defi}
 \label{defiAWIPi}
The absence of weak immediate profit  (AWIP) condition holds true if for all $\in \{0,\ldots,T\}$
$$\overline{\mathcal{R}_t^T} \cap L^0(\R_+,\cF_t)=\{0\},$$  where the closure of $\mathcal{R}_t^T$ is taken with respect to the convergence in probability.
 \end{defi}
We will see in Lemma \ref{lemstrict} that the AIP condition is not necessarily equivalent to  AWIP.
Before, in the case $d=1$ we show that  AWIP  may be equivalent to AIP condition under an extra closeness condition. It also provides a characterization through (absolutely continuous) martingale measures.
\begin{theo}\label{theo-EqWNFL} Assume that the case $d=1$. The following statements are equivalent:
\begin{enumerate}
\item AWIP  holds.
\item For every $t \in \{0,\ldots,T\}$, there exists $Q\ll P$ with $\E(dQ/dP|\cF_t)=1$ such that $(S_u)_{u\in \{t,\ldots, T\}}$ is a $Q$-martingale.
\item AIP holds and $\overline{\mathcal{R}_t^T} \cap L^0(\R,\cF_t)=\mathcal{R}_t^T \cap L^0(\R,\cF_t)$ for every $t \in \{0,\ldots,T\}.$
\end{enumerate}
\end{theo}
The proof is based on classical Hahn-Banach Theorem arguments, see for example the textbooks of \cite{DelSch05} and  \cite{KS}.
\begin{rem}\label{propEqWNFL-AIP}
From above, it is clear that AIP and AWIP are equivalent if  $\mathcal{R}_t^T$ is  closed. Therefore, we deduce by Lemma \ref{lemstrict}  that $\mathcal{R}_t^T$ is  not necessarily closed under AIP. \\
Suppose now that $P(\essinf[\cF_t]S_{t+1}=S_t)=P(\esssup[\cF_t]S_{t+1}=S_t)=0$ for all $t\in \{0\ldots,T-1\}$. Then, using Lemma \ref{NAAIP},
 AIP is equivalent to NA. Under NA, the set $\mathcal{R}_{t}^T$ is closed in probability for every $t\in \{0\ldots,T-1\}$ and 
 Theorem \ref{theo-EqWNFL} implies that AWIP, AIP and NA are equivalent conditions.\\
\end{rem}
{\sl Proof.} First we prove that 1. implies 2. Suppose that AWIP  holds and fix some $t\in \{0,\ldots, T\}$. We may suppose without loss of generality that the process  $S$ is integrable under $P$. Under AWIP, we then have $\overline{\mathcal{R}_t^T} \cap L^1(\R_+,\cF_t)=\{0\}$ where the closure is taken in $L^1$. Therefore,   for every nonzero $x\in L^1(\R_+,\cF_t)$, there exists by the Hahn-Banach theorem a non-zero $Z_x\in L^{\infty}(\R_+,\cF_T)$ such that (recall that $\mathcal{R}_t^T$ is a cone) $\E Z_xx>0$ and  $\E Z_x\xi\le 0$ for every $\xi\in  \mathcal{R}_t^T$. Since $-L^1(\R_+,\cF_T)\subseteq \mathcal{R}_t^T$, we deduce that $Z_x\ge 0$ and we way renormalise $Z_x$ so that $\| Z_x\|_{\infty}=1$.  Let us consider the family
$$\cG=\{\{ \E(Z_x|\cF_t)>0\},~x\in L^1(\R_+,\cF_t)\setminus \{0\}\}.$$ Consider any non-null set $\Gamma\in \cF_t$. Taking $x=1_{\Gamma}\in L^1(\R_+,\cF_t)\setminus \{0\}$, since $\E(Z_x1_{\Gamma})>0$, we deduce that $\Gamma$ has a non-null intersection with $\{\E(Z_x|\cF_t)>0\}$. By \cite[Lemma 2.1.3]{KS}, we deduce an at most  countable subfamily $(x_i)_{i\ge 1}$ such that
the union $\bigcup_i\{\E(Z_{x_i}|\cF_t)>0\}$ is of full measure. Therefore, $$Z=\sum_{i=1}^{\infty}2^{-i}Z_{x_i}\ge 0$$ is such that $\E(Z|\cF_t)>0$ and we define $Q\ll P$ such that $dQ=(Z/\E(Z|\cF_t))dP$. As the subset $\{ \sum_{u=t+1}^T \theta_{u-1}\Delta S_u,~\theta_{u-1}\in L(\R,\cF_{u-1}) \}$ is a linear vector space contained in $\mathcal{R}_t^T$, we deduce that $(S_u)_{u\in \{t,\ldots, T\}}$ is a $Q$-martingale.

We now prove that 2. implies 3.
Suppose that for every $t\in \{0,\ldots, T\}$, there exists $Q\ll P$ such that $(S_u)_{u=t,\ldots, T}$ is a $Q$-martingale with  $\E(dQ/dP|\cF_t)=1$. Let us define for $u\in \{t,\ldots, T\}$, $\rho_u=\E_{\bP}(dQ/d\bP|\cF_u)$ then $\rho_u \geq 0$ and $\rho_t=1$. Consider $\gamma_t\in \mathcal{R}_t^T \cap L^0(\R_+,\cF_t)$, i.e. $\gamma_t$ is $\cF_t$-measurable and is of the form $\gamma_t=\sum_{u=t}^{T-1} \theta_{u}\Delta S_{u+1}-\epsilon_T^+$.
Since $\theta_u$ is $\cF_u$-measurable, $\theta_{u}\Delta S_{u+1}$ admits a generalized conditional expectation under $Q$ knowing $\cF_u$ and we have by assumption that $\E_Q(\theta_{u}\Delta S_{u+1}|\cF_u)=0$. The tower law implies that a.s.
$$\gamma_t=\E_Q(\gamma_t|\cF_t)=\sum_{u=t}^{T-1} \E_Q(\E_Q(\theta_{u}\Delta S_{u+1}|\cF_u)|\cF_t)-\E_Q(\epsilon_T^+|\cF_t)=-\E_Q(\epsilon_T^+|\cF_t).$$
Hence $\gamma_t=0$ a.s., i.e. AIP holds. It remains to show that $\overline{\mathcal{R}_t^T} \cap L^0(\R,\cF_t)\subseteq \mathcal{R}_t^T \cap L^0(\R,\cF_t)$.

Consider first a one step model, where $(S_u)_{u\in \{T-1, T\}}$ is a $Q$-martingale with  $\rho_T\geq 0$ and $\rho_{T-1}=1$.
Suppose that $\gamma^n=\theta_{T-1}^n\Delta S_T-\epsilon_T^{n+}\in L^0(\R,\cF_{T-1})$ converges in probability  to $\gamma^{\infty}\in L^0(\R,\cF_{T-1})$. We need to show that $\gamma^{\infty}\in \mathcal{R}_{T-1}^T \cap L^0(\R,\cF_{T-1})$.

On the $\cF_{T-1}$-measurable set $\Lambda_{T-1}:=\{\liminf_n|\theta_{T-1}^n|<\infty\}$, by \cite[Lemma 2.1.2]{KS}, we may assume w.l.o.g. that $\theta_{T-1}^n$ is convergent to some $\theta_{T-1}^{\infty}$ hence $\epsilon_T^{n+}$ is also convergent and we can  conclude that $\gamma^{\infty}1_{\Lambda_{T-1}}\in \mathcal{R}_{T-1}^T \cap L^0(\R,\cF_{T-1})$.

Otherwise, on $\Omega\setminus \Lambda_{T-1}$, we use the normalized sequences for $i \in \{1,\ldots,d\}$
$$\tilde \theta_{T-1}^{n,i}:=\theta_{T-1}^{n,i}/(|\theta_{T-1}^n|+1),  \;\; \tilde\epsilon_T^{n+}:=\epsilon_T^{n+}/(|\theta_{T-1}^n|+1).$$ By \cite[Lemma 2.1.2]{KS} again, we may assume taking $d+1$ sub-sequences that a.s. $\tilde \theta_{T-1}^n\to \tilde \theta_{T-1}^{\infty}$, $\tilde \epsilon_T^{n+}\to \tilde \epsilon_T^{\infty+}$ and
$$\tilde \theta_{T-1}^{\infty}\Delta S_T-\tilde \epsilon_T^{\infty+}=0 \mbox{ a.s.}$$
Remark that $|\tilde \theta_{T-1}^{\infty}|=1$ a.s. First consider the subset $\Lambda_{T-1}^2:=\left(\Omega\setminus \Lambda_{T-1}\right)\cap \{\tilde \theta_{T-1}^{\infty}=1\} \in \cF_{T-1}$ on which $\Delta S_T\ge 0$ a.s. Since  $\E_{Q}(\Delta S_T1_{\Lambda_{T-1}^2}|\cF_{T-1})=0$ a.s., we get
that
$\rho_T \Delta S_T1_{\Lambda_{T-1}^2}=0$ a.s. Hence  $\rho_T \gamma^n1_{\Lambda_{T-1}^2}=-\rho_T\epsilon_T^{n+}1_{\Lambda_{T-1}^2} \leq 0$ a.s. Taking the limit, we get that $\rho_T \gamma^{\infty}1_{\Lambda_{T-1}^2}\le 0$ a.s. and, since $\gamma^{\infty}\in L^0(\R,\cF_{T-1})$, we deduce that $\rho_{T-1}\gamma^{\infty}1_{\Lambda_{T-1}^2}\le 0$ a.s. Recall that  $\rho_{T-1}=1$ hence $\gamma^{\infty}1_{\Lambda_{T-1}^2}\le 0$ a.s. and $\gamma^{\infty}1_{\Lambda_{T-1}^2}\in \mathcal{R}_{T-1}^T \cap L^0(\R,\cF_{T-1})$. On the subset $\left(\Omega\setminus \Lambda_{T-1}\right)\cap \{\tilde \theta_{T-1}^{\infty}=-1\}$ we may argue similarly and the conclusion follows in the one step model.

We now show the result in multi-step models by recursion. Fix some $s \in \{t,\ldots,T-1\}$. We show that $\overline{\mathcal{R}_{s+1}^T} \cap L^0(\R,\cF_{s+1})\subseteq \mathcal{R}_{s+1}^T \cap L^0(\R,\cF_{s+1})$ implies the same property for $s$ instead of $s+1$. By assumption $(S_u)_{u\in \{s,\ldots, T\}}$ is a $Q$-martingale with  $\E_{P}(dQ/dP|\cF_u)=\rho_u \geq 0$ for  $u\in \{s,\ldots, T\}$ and $\rho_s=1$. Suppose that
$$\gamma^n=\sum_{u=s+1}^{T} \theta_{u-1}^n\Delta S_{u}-\epsilon_T^{n+} \in \mathcal{R}_{s}^T \cap L^0(\R,\cF_s) \mbox{ converges to } \gamma^{\infty}\in L^0(\R,\cF_{s}).$$
If $\gamma^{\infty}=0$ there is nothing to prove.
As before on the $\cF_{s}$-measurable set $\Lambda_{s}:=\{\liminf_n|\theta_{s}^n|<\infty\}$,  we may assume w.l.o.g. that $\theta_{s}^n$  converges to $\theta_{s}^{\infty}$. Therefore on $\Lambda_{s}$
$$\sum_{u=s+2}^{T} \theta_{u-1}^n\Delta S_{u}-\epsilon_T^{n+}
=\gamma^n - \theta_{s}^n\Delta S_{s+1} \to
\gamma^{\infty} - \theta_{s}^{\infty} \Delta S_{s+1}$$
and by the induction hypothesis,  $\sum_{u=s+2}^{T} \theta_{u-1}^n\Delta S_{u}-\epsilon_T^{n+}$ also converges to an element of $\mathcal{R}_{s+1}^T\cap L^0(\R,\cF_{s+1})$ and we  conclude that $\gamma^{\infty} 1_{\Lambda_{s}} \in  \mathcal{R}_{s}^T \cap  L^0(\R,\cF_{s})$.\\
On $\Omega\setminus \Lambda_{s-1}$, we use the normalisation procedure as before, and deduce the equality
$$\sum_{u=s+1}^{T} \tilde \theta_{u-1}^{\infty}\Delta S_{u}-\tilde \epsilon_T^{{\infty}+}=0 \mbox{ a.s.}$$
for some $\tilde \theta_{u}^{\infty}\in L^0(\R,\cF_{u})$, $u \in \{s,\ldots,T-1\}$ and $\tilde \epsilon_T^{{\infty}+}\ge 0$ such that $|\tilde \theta_{s}^{\infty}|=1$ a.s. We then argue on $\Lambda_{s}^2:=\left(\Omega\setminus \Lambda_{s-1}\right)\cap \{\tilde \theta_{s}^{\infty}=1\}\in \cF_{s}$ and $\Lambda_{s}^3:=\left(\Omega\setminus \Lambda_{s-1}\right)\cap \{\tilde \theta_{s}^{\infty}=-1\}\in \cF_{s}$ respectively. \\
When $\tilde \theta_{s}^{\infty}=1$, we deduce that
$$\Delta S_{s+1}+\sum_{u=s+2}^{T} \tilde \theta_{u-1}^{\infty}\Delta S_{u} -\tilde \epsilon_T^{{\infty}+}=0 \mbox{ a.s., i.e. } \Delta S_{s+1}\in \cP_{s+1,T}(0)$$ hence $\Delta S_{s+1}\ge \pi_{s+1,T}(0)=0$ a.s. under AIP, see Theorem \ref{thoNip}. \\Since  $\E_{Q}(\Delta S_{s+1}1_{\Lambda_{s}^2}|\cF_{s})=0$ a.s.,   $\rho_{s+1}\Delta S_{s+1}1_{\Lambda_{s}^2}=0$ a.s.
So,
$$\rho_{s+1}\gamma^n 1_{\Lambda_{s}^2}=\sum_{u=s+2}^{T} \theta_{u-1}^n \rho_{s+1} 1_{\Lambda_{s}^2}\Delta S_{u}-\epsilon_T^{n+} \rho_{s+1} 1_{\Lambda_{s}^2}\in \mathcal{R}_{s+1}^T \cap L^0(\R,\cF_{s+1}).$$
Hence $\rho_{s+1}\gamma^{\infty} 1_{\Lambda_{s}^2} \in \mathcal{R}_{s+1}^T \cap L^0(\R,\cF_{s+1})$ by induction.
As $\rho_{s+1}\gamma^{\infty}1_{\Lambda_{s}^2}$ admits a generalized conditional expectation knowing $\cF_s$,
we deduce by the tower law that a.s.
\begin{eqnarray*}
1_{\Lambda_{s}^2} \E(\rho_{s}\gamma^{\infty}|\cF_s) & = &
\E(\rho_{s+1}\gamma^{\infty}1_{\Lambda_{s}^2}|\cF_s) \\
& = &
\sum_{u=s+2}^{T} 1_{\Lambda_{s}^2} \E\left(\theta_{u-1}^{\infty}\E  \left(\frac{dQ}{dP}\Delta S_{u}|\cF_{u-1}\right) |\cF_s\right)
-1_{\Lambda_{s}^2} \E(\epsilon_T^{\infty+} \rho_{s+1} |\cF_s) \\
& \leq & 0,
\end{eqnarray*}
since $(S_u)_{u\in \{s,\ldots, T\}}$ is a $Q$-martingale.
Hence $\rho_{s}\gamma^{\infty}1_{\Lambda_{s}^2}\le 0$ a.s.  As  $\rho_{s}=1,$ $\gamma^{\infty}1_{\Lambda_{s}^2}\le 0$ a.s. so that $\gamma^{\infty}1_{\Lambda_{s}^2}\in \mathcal{R}_{s}^T \cap L^0(\R,\cF_{s})$.

 Finally, notice that the AIP condition implies AWIP  as soon as the equality $\overline{\mathcal{R}_t^T} \cap L^0(\R_+,\cF_t)=\mathcal{R}_t^T \cap L^0(\R_+,\cF_t)$ holds for every $t \in \{0,\ldots,T\}$.   \fdem

\begin{lemm} \label{lemstrict} The AIP condition is not necessarily equivalent to AWIP.
\end{lemm}
{\sl Proof.}
Assume that $d=1$.  Let us consider a positive process $(\tilde S_t)_{t\in \{0,\ldots,T\}}$ which is a $P$-martingale. We suppose that $\essinf[\cF_0]\tilde S_1<\tilde S_1$ a.s., which holds in particular
if  $\tilde S$ a geometric Brownian motion as $\essinf[\cF_0]\tilde S_1=0$ a.s.  Let us define $S_t:=\tilde S_t$ for $t\in\{1,\ldots,T\}$ and $S_0:=\essinf[\cF_0]S_1$. We have $\essinf[\cF_0]S_1\le S_0$ and $\esssup[\cF_0]S_1\ge \essinf[\cF_0]S_1=S_0$ hence AIP holds at time $0$ (see Remark \ref{foi}). Moreover, by the martingale property (see Theorem \ref{theo-EqWNFL}), AIP and also AWIP hold at any time $t\in\{1,\ldots,T\}$. Let us suppose that AWIP  holds true at $t=0$. Using Theorem \ref{theo-EqWNFL}, there exists $\rho_T\ge 0$ with $\E(\rho_T)=1$ such that $S$ is a $Q$-martingale where $dQ=\rho_TdP$. Therefore, $\E(\rho_T\Delta S_1)=0$. Since $\Delta S_1>0$ by assumption, we deduce that $\rho_T=0$ hence a contradiction. \fdem

\section{Explicit pricing of a convex payoff under AIP}
\label{seexpli}
The aim of this section is to obtain some results in a particular model where $d=1$, $\essinf[\cF_{t-1}]S_t=k^d_{t-1}S_{t-1}$ a.s. and  $\esssup[\cF_{t-1}]S_t=k^u_{t-1}S_{t-1}$ a.s. for every $t\in \{1,\ldots,T\}$ with $(k^d_{t-1})_{t\in \{1,\ldots,T\}}$,  $(k^u_{t-1})_{t\in \{1,\ldots,T\}}$ and $S_0$  are {\it deterministic} non-negative numbers. We obtain the same computative scheme (see \eqref{functional scheme}) as in \cite{CV} but  assuming only AIP and not NA. We also propose some numerical experiments.
\subsection{The algorithm}

 \begin{theo}\label{MT1} Suppose that the model is defined by $\essinf[\cF_{t-1}]S_t=k^d_{t-1}S_{t-1}$ {\rm a.s.} and $\esssup[\cF_{t-1}]S_t=k^u_{t-1}S_{t-1}$ {\rm a.s.} where $(k^d_{t-1})_{t\in \{1,\ldots,T\}}$, $(k^u_{t-1})_{t\in \{1,\ldots,T\}}$ and $S_0$ are deterministic non-negative numbers.
 \begin{itemize}
\item The AIP condition holds true if and only
if  $k^d_{t-1}\in [0, 1]$ and $k^u_{t-1}\in [1, +\infty]$ for all $t\in \{1,\ldots,T\}$.

\item Suppose that the AIP condition holds. Let $h:\R \to \R$ be a non-negative convex function with ${\rm dom \,h }=\R$ such that $\lim_{z\to +\infty}\frac{h(z)}{z}\in [0,\infty)$. Then the infimum super-hedging cost of the European contingent claim $h(S_T)$ is a price and it is given by 
\begin{equation}\label{functional scheme}
\begin{split}
\pi_{t,T}(h) & =h(t,S_t)\in \mathcal{P}_{t,T}(h(S_T)) \mbox{a.s.}\\
h(T,x) & =  h(x) \\
h(t-1,x) & = \lambda_{t-1}h \left(t,k^d_{t-1}x\right)+(1-\lambda_{t-1})h\left(t,k^u_{t-1}x\right),
\end{split}
\end{equation}
where $\lambda_{t-1}= \frac{k^u_{t-1}-1}{k^u_{t-1}-k^d_{t-1}}\in [0,1]$ and
$1-\lambda_{t-1}= \frac{1-k^d_{t-1}}{k^u_{t-1}-k^d_{t-1}}\in [0,1]$,
with the following conventions. When $k^d_{t-1}=k^u_{t-1}=1$ or $S_{t-1}=0$,
 $\lambda_{t-1}=\frac{0}{0}=0$ and $1-\lambda_{t-1}=1$
and
when $k^d_{t-1}<k^u_{t-1}=\infty,$
\begin{small}
\begin{equation}\label{functional scheme-convention}
\begin{split}
\lambda_{t-1}  = &\frac{\infty}{\infty}=1\\
(1-\lambda_{t-1})h(t,(+\infty)x)  = & (1-k^d_{t-1})x\frac{h(t,(+\infty x))}{(+\infty x)}\\
=  & (1-k^d_{t-1}) x\lim_{z\to +\infty}\frac{h(z)}{z}.
\end{split}
\end{equation}
\end{small}
\end{itemize}
Moreover, for every $t\in \{1,\ldots,T\}$, $\lim_{z\to +\infty}\frac{h(z)}{z}=\lim_{z\to +\infty}\frac{h(t,z)}{z}$ and $ h(\cdot,x)$ is non-increasing for all $x\geq 0$.
\end{theo}
In the proof, the strategy associated to the infimum super-hedging  price is given and, this result is illustrated through a numerical experiment  in Section \ref{secnum}. \smallskip

{\sl Proof.}
The  conditions $k^d_{t-1}\in [0, 1]$ and $k^u_{t-1}\in [1, +\infty]$ for all $t\in \{1,\ldots,T\}$ are equivalent to the AIP condition (see Remark \ref{foi}). We denote $M=\frac{h(\infty)}{\infty}$ and $M_t=\lim_{z\to +\infty}\frac{h(t,z)}{z}$. We prove the second statement. Assume that AIP holds true.  We establish (i) the recursive formulation $\pi_{t,T}(h(S_T))=h(t,S_t)$ given by \eqref{functional scheme}, (ii) $h(t,\cdot) \geq h(t+1,\cdot)$ and (iii) $M_t=M_{t+1}$. The case $t=T$ is immediate. As $h:\R \to \R$ is a convex function with ${\rm dom \, h}=\R$, $h$ is clearly a $\cF_{T-1}$-normal integrand, we can apply Corollary \ref{fenchngdconv} (see \eqref{PricingFormula} and \eqref{theta}) and we get that a.s.
\begin{equation}\label{eqfirststep}
\begin{split}
\pi_{T-1,T}(h(S_T))& =  h(k^d_{T-1}S_{T-1} ) + \theta_{T-1}^*\left( S_{T-1}-k^d_{T-1}S_{T-1} \right),\\
\theta_{T-1}^* & =  \frac{h(k^u_{T-1}S_{T-1})-h(k^d_{T-1}S_{T-1})}{k^u_{T-1}S_{T-1} -k^d_{T-1}S_{T-1}},
\end{split}
\end{equation}
where we use the conventions $\theta_{T-1}^*=\frac{0}{0}=0$ if either $S_{T-1}=0$ or $k^u_{T-1}=k^d_{T-1}=1$ and $\theta_{T-1}^*=\frac{h(\infty)}{\infty}=M$ if $k^d_{T-1}<k^u_{T-1}=+\infty$. Moreover, using \eqref{lavieestbelle}, we obtain that
$$\pi_{T-1,T}(h(S_T))+ \theta_{T-1}^* \Delta S_T \geq h(S_T)\mbox{ a.s. i.e. } \pi_{T-1,T}(h(S_T))\in \cP_{T-1,T}(h(S_T)).$$ So, using Lemma \ref{lemouf}, we get that
$\cP_{T-2,T}(h(S_T))=\cP_{T-2,T-1}(\pi_{T-1,T}(h(S_T)))$,
$$\pi_{T-2,T}(h(S_T))=\pi_{T-2,T-1}(\pi_{T-1,T}(h(S_T)))$$ and we may continue the recursion as soon as $\pi_{T-1,T}(h(S_T))=h(T-1,S_{T-1})$ where  $h(T-1,\cdot)$ satisfies \eqref{functional scheme}, is convex with domain equal to $\R$, is such that $h(T-1,z)\geq 0$ for all $z \geq 0$ and $M_{T-1}=M\in  [0,\infty)$. To see that we distinguish three cases. If either $S_{T-1}=0$ or $k^u_{T-1}=k^d_{T-1}=1$,
$\pi_{T-1,T}(h(S_T))=h(S_{T-1})$ and $h(T-1,z)=h(z)=h(T,z)$ satisfies all the required conditions. If  $k^d_{T-1}<k^u_{T-1}=+\infty$,
$$\pi_{T-1,T}(h(S_T))=h(k^d_{T-1}S_{T-1} ) + M\left( S_{T-1}-k^d_{T-1}S_{T-1} \right)=h(T-1,S_{T-1})$$ with
\bean h(T-1,z) & =& h(k^d_{T-1}z ) + M z\left(1 -k^d_{T-1} \right)\\
& =& \lim_{k^u\to +\infty}\left( \frac{k^u-1}{k^u-k^d_{T-1}}h(k^d_{T-1}z)+\frac{1-k^d_{T-1}}{k^u-k^d_{T-1}}h(k^uz)\right),
\eean
using \eqref{functional scheme-convention}. The term in the r.h.s. above is larger than $h(z)=h(T,z)$ by convexity since
$$\frac{k^u-1}{k^u-k^d_{T-1}}k^d_{T-1}z + \frac{1-k^d_{T-1}}{k^u-k^d_{T-1}}k^uz=z.$$ As $k^d_{T-1}\in [0, 1]$ and $M\in  [0,\infty)$, $h(T-1,z) \geq 0$  for all $z \geq 0$, we get that $h(T-1,\cdot)$ is convex function with domain equal to $\R$ since $h$ is so. The function $h(T-1,\cdot)$  also satisfies \eqref{functional scheme} (see \eqref{functional scheme-convention}).
Finally $$M_{T-1}=\lim_{z \to + \infty} k^d_{T-1}\frac{h(k^d_{T-1}z )}{k^d_{T-1}z} + M \left(1 -k^d_{T-1} \right)=M.$$

The last case is when  $S_{T-1}\neq0$ and $k^u_{T-1}\neq k^d_{T-1}$ and $k^u_{T-1}<+\infty$. It is clear that \eqref{eqfirststep} implies \eqref{functional scheme}.
Moreover as $k^d_{T-1}\in [0, 1]$ and $k^u_{T-1}\in [1, +\infty)$, 
$$\lambda_{T-1}= \frac{k^u_{T-1}-1}{k^u_{T-1}-k^d_{T-1}}\in [0,1] \mbox{ and }1-\lambda_{T-1}= \frac{1-k^d_{T-1}}{k^u_{T-1}-k^d_{T-1}}\in [0,1]$$ and  \eqref{functional scheme}  implies that $h(T-1,z) \geq 0$  for all $z \geq 0$,
 $h(T-1,\cdot)$ is convex with domain equal to $\R$ since $h$ is so. Moreover,
 \begin{small}
 $$M_{T-1}= \lambda_{T-1} k^d_{T-1} \lim_{z \to + \infty}\frac{ h(k^d_{T-1}z)}{k^d_{T-1}z}+(1-\lambda_{T-1}) k^u_{T-1} \lim_{z \to + \infty}\frac{ h(k^u_{T-1}z)}{k^u_{T-1}z}= M,$$
 \end{small}
 since
$$\lambda_{T-1}k^d_{T-1}+(1-\lambda_{T-1})k^u_{T-1}=1.$$


\fdem

\begin{rem} The infimum super-hedging cost of the European contingent claim $h(S_T)$ in our  model is a price, precisely the same than the price we get in a binomial model   $S_{t}\in \{k^d_{t-1}S_{t-1},k^u_{t-1}S_{t-1}\}$ a.s., $t\in \{1,\ldots,T\}$. Moreover, as in Corollary \ref{aipcall}, one can prove that the AIP condition holds at every instant $t$ if and only
if the super-hedging prices of some European call option at $t$ are non-negative. The advantage of this generalization is to provide a statistical principle to compute the minimal price and the super-hedging strategy, see next section. 

\end{rem}

\subsection{Numerical experiments}
\label{secnum}
\subsubsection{Calibration}

In this section, we suppose that the discrete dates are given by  $t_i^n=\frac{iT}{n}$, $i\in \{0,\ldots,n\}$ where $n\ge 1$. We assume that  $$k_{t_{i-1}^n}^u=1+\sigma_{t_{i-1}^n}\sqrt{\Delta t_i^n} \mbox{ and }k_{t_{i-1}^n}^d=1-\sigma_{t_{i-1}^n}\sqrt{\Delta t_i^n}\ge 0,$$ where $t\mapsto \sigma_t$ is a positive Lipschitz-continuous function on $[0,T]$.
Note that the assumptions on the multipliers $k_{t_{i-1}^n}^u$ and $k_{t_{i-1}^n}^d$ imply that
\bea \label{Ineq} \left |\frac{S_{t_{i+1}^n}}{S_{t_i^n}}-1 \right|\le \sigma_{t^n_i}\sqrt{\Delta t_{i+1}^n,}{\rm \,a.s.}\eea
By Theorem \ref{MT1}, we deduce that the infimum super-hedging cost of the European Call option $(S_T-K)^+$ is given by $h^n\left(t_i^n,S_{t_i^n}\right)$ defined by (\ref{functional scheme}) with terminal condition $h^n(T,x)=g(x)=(x-K)^+$.  We extend the function $h^n$ on $[0,T]$ in such a way that $h^n$ is constant on each interval $[t_i^n,t_{i+1}^n[$, $i\in \{0,\ldots,n\}$. Such a scheme is proposed by Milstein \cite{Mil} where a convergence theorem is proved when the terminal condition, i.e. the payoff function is smooth. Precisely,   the sequence of functions $(h^n(t,x))_{n\geq 1}$  converges uniformly to $h(t,x)$,  solution  to the diffusion equation:
\begin{eqnarray}
\label{BS}
&&\partial_t h(t,x)+\sigma_t^2\frac{x^2}{2}\partial_{xx}h(t,x)=0,\quad h(T,x)=g(x).
\end{eqnarray}
In \cite{Mil}, it is supposed that  the successive derivatives of the solution of the P.D.E. solution $h$   are uniformly bounded. This is not the case for the Call payoff function $g$. On the contrary the successive derivatives of the solution of the P.D.E. explode at the horizon date, see \cite{LT}. In  \cite{BL}, it is proven that the uniform convergence still holds when the payoff function is not smooth provided that the successive derivatives of the solution of the P.D.E. do not explode {\sl too much}.

\begin{figure}[h!]
    \center
	\includegraphics[scale=0.75]{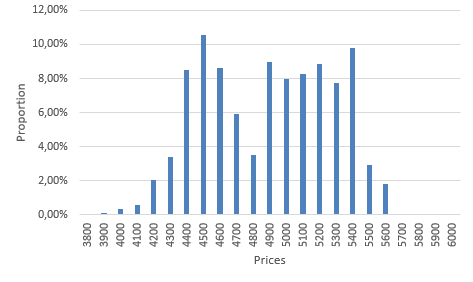}
    \caption{Distribution of the observed prices.}
    \label{Distri-price}
\end{figure}

Supposing that  $\Delta t_i^n$ is closed to $0$, we can identify  the observed prices of the Call option with the  theoretical prices limit $h(t,S_t)$ at any instant $t$, given by \eqref{BS}, to deduce an evaluation of  the deterministic function $t\mapsto \sigma_t$ and test  (\ref{Ineq}) on real data. The data set is composed of historical values of the french index CAC 40 and European call option prices of maturity $3$ months from the 23rd of October 2017 to the 19th of January 2018. The observed values of $S$ are distributed as in Figure \ref{Distri-price}.

For several strikes, matching the observed prices to the theoretical ones derived from the Black and Scholes formula with time-dependent volatility (see \eqref{BS}), we deduce the associated {\sl implied} volatility $t\mapsto \sigma_t$  and we compute the proportion of observations satisfying (\ref{Ineq}):

\begin{figure}[H]
    \center
	\includegraphics[scale=0.75]{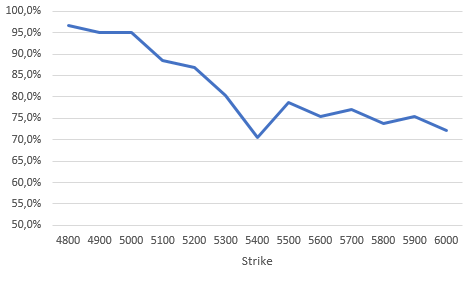}
    \caption{Ratio of observations satisfying (\ref{Ineq}) as a function of the strike.}
    \label{proportion}
\end{figure}

\begin{figure}[h!]
  \begin{center}
{\tiny
   \begin{tabular}{|c|c|c|c|c|c|c|c|c|c|c|c|c|c|}
   \hline
   Strike &  4800  & 4900 & 5000 & 5100 & 5200 & 5300 & 5400 & 5500 & 5600 & 5700 & 5800 & 5900 & 6000\\
\hline
Ratio 	&	96,7\% & 95,1\% & 95,1\% & 88,5\% & 86,9\% & 80,3\% & 70,5\% & 78,7\% & 75,4\% & 77,0\% & 73,8\% & 75,4\% & 72,1\%\\
\hline
\end{tabular}
     }
 \end{center}
 \end{figure}
The results are satisfactory for strikes lower that 5100. Note that when the strike increases  less prices's data are available for the Call option as the strike is too large with respect to the current price $S$, see Figure \ref{Distri-price}. This could explain the degradation of our results.

\subsubsection{Super-hedging prices}
We test the infimum super-hedging cost deduced from Theorem \ref{MT1} on some data set composed of historical daily closing values of the french index CAC 40 from the 5th of January 2015 to the 12th of March 2018. The interval $[0,T]$ we choose corresponds to one week composed of $5$ working days so that the discrete dates are $t^4_i$, $i\in \{0,\cdots,4 \}$ and $n=4$.  We first evaluate $\sigma_{t_i}$,  $i\in \{0,\cdots,3\}$ as
\bea \label{vol-max}
\sigma_{t_i}\:=\overline{\max}\left(\left |\frac{S_{t_{i+1}}}{S_{t_i}}-1 \right|/ \sqrt{\Delta t_{i+1}^4,}\right)\quad i\in \{0,\cdots,3\},
\eea
where $\overline{\max}$ is  the empirical maximum  taken over a one year sliding sample window of $52$ weeks. Notice that this estimation is model free and does not depend on the strike as it was the case in the preceding sub-section. So we estimate the volatility on 52 weeks and we implement our hedging strategy on the fifty third one. We then repeat the procedure by sliding the window of one week,
We observe the empirical average of the stock price  $S_{0}$ is equal to $4044$.

For a payoff function $g(x)=(x-K)^+$, we implement the strategy associated to the super-hedging cost given by Theorem \ref{MT1}. The super-hedging cost is given by $h(0,S_0)$ and, using \eqref{eqfirststep}, we compute  the super-hedging strategies $(\theta_{t_{i}^4}^*)_{i\in \{0,\ldots,3\}}$. We denote by $V_T$ the terminal value of our strategy starting from the minimal price $V_0=\pi_{0,T}=h(0,S_0)$:
$$V_T=V_0+\sum_{i=0}^3  \theta_{t_{i}^4}^* \Delta S_{ t_{i+1}^4}.$$
We study below the super-hedging error $\varepsilon_T=V_T-(S_T-K)^+$ for different strikes. 

{\bf Case where $K=4700$.} The distribution of the super-hedging error $\varepsilon_T$ for $K=4700$ is represented in  Figure \ref{distrErr}:

\begin{figure}[H]
    \center
	\includegraphics[scale=0.7]{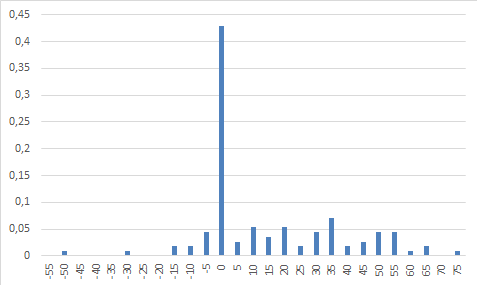}
    \caption{Distribution of the super-hedging error $\varepsilon_T=V_T-(S_T-K)^+$.}
    \label{distrErr}
\end{figure}

The empirical average of the error $\varepsilon_T$ is $12.76$ and its standard deviation is $21.65$. This result is rather satisfactory in comparison to the large value of the empirical mean of $S_{0}$ which is equal to $4844$. Notice that we observe $E(S_T-K)^+\simeq 282.69$. This empirically confirms the efficiency of our suggested method. The empirical probability of $\{\varepsilon_T<0\}$ is equal to $14.29\%$ but the Value at Risk at 95 \% is $-10.33$ which confirms that our strategy is conservative.

\begin{figure}[h!]
    \center
	\includegraphics[scale=0.7]{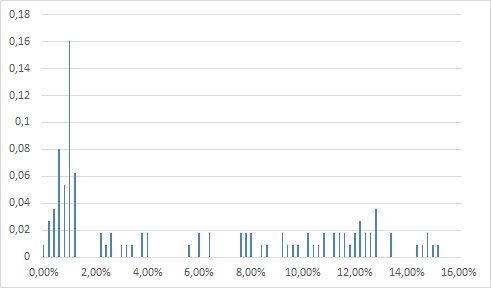}
    \caption{Distribution of the ratio $V_{0} / S_{0}$.}
    \label{distrPrixFinal}
\end{figure}

The empirical average of $V_0 / S_{0}$ is $5.61\%$ and its standard deviation is $5.14\%$. This is again satisfactory since the theoretical super-hedging price in incomplete market is often equal to $S_{0}$ (this is for example the case when $k^d=0$ and $k^u=\infty$, in particular when the dynamics of $S$ is modeled by a (discrete) geometric Brownian motion, see \cite{CGT}). Note that the loss of -50 is related to  so-called {\sl black friday} week that occures the  24th of June 2016. Large falls of risky assets were observed in European markets mainly explained by the Brexit vote. In particular, the CAC $40$ felt from $S_0=4340$ to $S_T=4106$, with a loss of  $-8\%$ on Friday.
\\

{\bf Case where $K=S_{0}$.}
We know present the ``at the money'' case. The empirical average of the error $\varepsilon_T=V_T-(S_T-K)^+$ is $35.69$ and its standard deviation is $34.11$. We observe $\E(S_{0})=4844$, $E(S_T-K)^+\simeq 38.15$, the probability $P(\varepsilon_T<0)=9.82\%$ and  the Value at Risk at 95 \% is $-11.41$. The empirical average of $V_0/S_0$ is $1.51\%$ and its standard deviation is $0.47\%$. \smallskip






{\bf Asymmetric case.} We now propose another estimation probably more natural of the parameters of the model:  
$k_{t_{j-1}^n}^d$ and $k_{t_{j-1}^n}^u$ are estimated as
$$k_{t_{i-1}^n}^d=\overline{\min} \frac{S_{t_{i}^n}}{S_{t_{i-1}^n}},\quad k_{t_{i-1}^n}^u=\overline{\max} \frac{S_{t_{i}^n}}{S_{t_{i-1}^n}},$$
where  the empirical minimum and maximum are taken over a one year sliding sample window of $52$ weeks, as previously.

{\bf Asymmetric case where $K=4700$.}\\
The distribution of the super-hedging error $\varepsilon_T$ for $K=4700$ is represented in  Figure \ref{distrErrAsym}:

\begin{figure}[H]
    \center
	\includegraphics[scale=0.7]{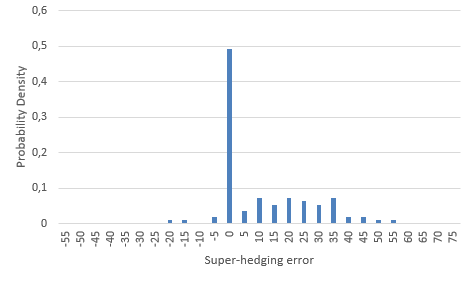}
    \caption{Distribution of the super-hedging error $\varepsilon_T=V_T-(S_T-K)^+$.}
    \label{distrErrAsym}
\end{figure}

The empirical average of the error $\varepsilon_T$ is $9.47$ and its standard deviation is $14.20$. This result is rather satisfactory in comparison to the large value of the empirical mean of $S_{0}$ which is equal to $4844$. The empirical probability of $\{\varepsilon_T<0\}$ is equal to $8.04\%$ and the Value at Risk at 95 \% is $-1.81$ which confirms that our strategy is conservative.

\begin{figure}[h!]
    \center
	\includegraphics[scale=0.7]{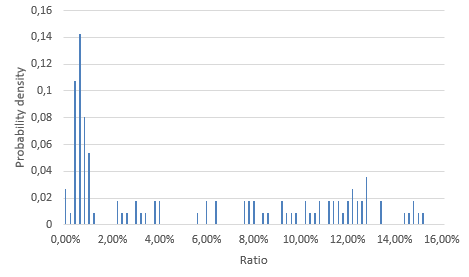}
    \caption{Distribution of the ratio $V_{0} / S_{0}$.}
    \label{distrPrixFinal}
\end{figure}

The empirical average of $V_0 / S_{0}$ is $5.52\%$ and its standard deviation is $5.22\%$.\\


{\bf Asymmetric case where $K=S_0$.}

The distribution of the super-hedging error $\varepsilon_T$ for $K=S_0$ is represented in  Figure \ref{distrErr2}:

\begin{figure}[H]
    \center
	\includegraphics[scale=0.7]{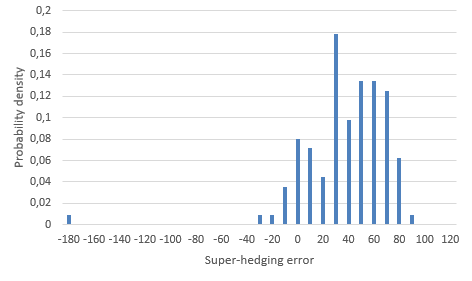}
    \caption{Distribution of the super-hedging error $\varepsilon_T$.}
    \label{distrErr2}
\end{figure}

The empirical average of the error $\varepsilon_T=V_T-(S_T-K)^+$ is $33.37$ and its standard deviation is $32.78$. The probability $P(\varepsilon_T<0)=12.5\%$. The Value at Risk at 95 \% is $-9.29.$ 

\begin{figure}[h!]
    \center
	\includegraphics[scale=0.7]{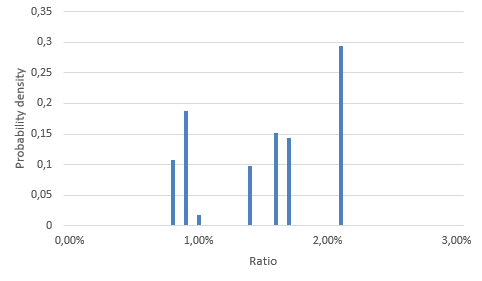}
    \caption{Distribution of the ratio $V_{0} / S_{0}$.}
    \label{distrPrixFinal2}
\end{figure}

The empirical average of $V_0 / S_{0}$ is $1.47\%$ and its standard deviation is $0.49\%$.\\

We now compare the result of both methods is the table below.
\begin{figure}[h!]
  \begin{center}
{\tiny
   \begin{tabular}{|c|c|c|c|c|c|c|}
   \hline
   & Mean of $V_{0} / S_{0}$ & Variance of $V_{0} / S_{0}$ & Mean of $\varepsilon_T$ & Variance of $\varepsilon_T$ &  $P(\varepsilon_T<0)$ &   VaR 95 \%\\
   \hline
Symmetric &  5.61\%  & 5.14 \%& 12.76 & 21.65 & 14.29 \%& -10.33 \\
\hline
Asymmetric	&	5.52\% & 5.22\% & 9.47 & 14.20 & 8.04\% & -1.81 \\
\hline
\end{tabular}
\caption{Comparison of the two methods of estimation for $K=4700$. The mean of $S_0$ is 4844,93 and the mean of $(S_0-K)^+$ is 278,73.}
     }
 \end{center}
 \end{figure}

\begin{figure}[h!]
  \begin{center}
{\tiny
   \begin{tabular}{|c|c|c|c|c|c|c|}
   \hline
   & Mean of $V_{0} / S_{0}$ & Variance of $V_{0} / S_{0}$ & Mean of $\varepsilon_T$ & Variance of $\varepsilon_T$ &  $P(\varepsilon_T<0)$ &   VaR 95 \%\\
   \hline
Symmetric &  1.51\%  & 0.47 \%& 35.69 & 34.11 & 9.82 \%& -11.41 \\
\hline
Asymmetric	&	1.47\% & 0.49\% & 33.37 & 32.78 & 12.50\% & -9.29 \\
\hline
\end{tabular}
\caption{Comparison of the two methods of estimation for $K=S_0$. The mean of $S_0$ is 4844,93.}
     }
 \end{center}
 \end{figure}
 The asymmetric method perform better than the symmetric one which is not a surprise.

\section{Appendix}
\label{secappendix}

{\sl Proof of Corollary \ref{lemma-essup-h(set)}.} For all $X \in \cX$, $\esssup[\cH] \{h(X), \, X \in \cX \} \geq h(X)$ a.s. and as $\esssup[\cH] \{h(X), \, X \in \cX \}$ is $\cH$-measurable,  we get that a.s. 
\begin{eqnarray*}
\esssup[\cH] \{h(X), \, X \in \cX \} &\geq &\esssup[\cH] h(X)\\
\esssup[\cH] \{h(X), \, X \in \cX \} & \geq  & \sup_{X \in \cX} \esssup[\cH] h(X).
\end{eqnarray*}
Conversely, for all $X \in \cX$, $\sup_{X \in \cX} \esssup[\cH] h(X)  \geq  h(X)$ a.s.

\noindent If $\sup_{X \in \cX} \esssup[\cH] h(X) $ is $\cH$-measurable, we may conclude
that a.s. $$\sup_{X \in \cX} \esssup[\cH] h(X)  \geq \esssup[\cH] \{h(X), \, X \in \cX \}.$$ Using Proposition \ref{lemma-essup-h(X)}, we get that a.s. 
\begin{eqnarray*}
\sup_{X \in \cX} \esssup[\cH] h(X) & =& \sup_{X \in \cX} \sup_{x \in \supp[\cH]X} h(x)=\sup_{x\in \cup_{X \in \cX}  \supp[\cH]X} h(x).
\end{eqnarray*}
Since $\cup_{X \in \cX}\supp[\cH]X$ is $\cH$-measurable and closed-valued, Lemma \ref{mesmes} implies that $\sup_{x\in \cup_{X \in \cX}  \supp[\cH]X} h(x)$ is $\cH$-measurable and the proof is complete.
\fdem


\end{document}